\documentclass[a4paper,fleqn,usenatbib]{mnras}

\usepackage[T1]{fontenc}
\usepackage{ae,aecompl}

\usepackage{graphicx}	
\usepackage{amsmath}	
\usepackage{amssymb}	
\usepackage[version=3]{mhchem}
\usepackage{subfig}

\title[A primordial origin for \ce{O2} in comets]{A primordial origin
  for molecular oxygen in comets: A chemical kinetics study of the
  formation and survival of O$_2$ ice from clouds to disks}

\author[V. Taquet et al.]{
V. Taquet,$^{1}$\thanks{E-mail: taquet@strw.leidenuniv.nl (VT)}
K. Furuya,$^{1}$
C. Walsh,$^{1}$
and E. F. van Dishoeck,$^{1,2}$
\\
$^{1}$Leiden Observatory, Leiden University, P.~O.~Box 9531, 2300~RA Leiden, The Netherlands\\
$^{2}$Max-Planck-Institut f\"{u}r extraterretrische Physik, Giessenbachstrasse 1, 85748 Garching, Germany
}

\date{Accepted XXX. Received YYY; in original form ZZZ}

\pubyear{2016}

\begin{document}
\label{firstpage}
\pagerange{\pageref{firstpage}--\pageref{lastpage}}
\maketitle

\begin{abstract}
Molecular oxygen has been confirmed as the fourth most abundant
molecule in cometary material (\ce{O2}/\ce{H2O} $\sim 4$ \%) and is
thought to have a primordial nature, i.e., coming from the
interstellar cloud from which our solar system was formed. However,
interstellar \ce{O2} gas is notoriously difficult to detect and has
only been observed in one potential precursor of a solar-like
system. Here, the chemical and physical origin of \ce{O2} in comets is
investigated using sophisticated astrochemical models. Three origins
are considered: i) in dark clouds, ii) during forming protostellar
disks, and iii) during luminosity outbursts in disks.  
The dark cloud models show that reproduction of the observed abundance
of \ce{O2} and related species in comet 67P/C-G requires a low H/O
ratio facilitated by a high total density ($\geq 10^5$ cm$^{-3}$), and
a moderate cosmic ray ionisation rate ($\leq 10^{-16}$ s$^{-1}$) while
a temperature of 20 K, slightly higher than the typical temperatures
found in dark clouds, also enhances the production of \ce{O2}. 
Disk models show that \ce{O2} can only be formed in the gas phase in
intermediate disk layers, and cannot explain the strong correlation
between \ce{O2} and \ce{H2O} in comet 67P/C-G together with the weak
correlation between other volatiles and \ce{H2O}. However, primordial
\ce{O2} ice can survive transport into the comet-forming regions of
disks.  
Taken together, these models favour a dark cloud (or "primordial'')
origin for \ce{O2} in comets, albeit for dark clouds which are warmer and
denser than those usually considered as solar system progenitors.  
\end{abstract}




\section{Introduction}
\label{introduction}

Molecular oxygen, \ce{O2}, is a dominant component of Earth's 
atmosphere (21\% by volume).  
Because it is a byproduct of photosynthesis (and also a reactant in 
cellular respiration), it is considered as a potential marker for 
biological activity on terrestrial-like exoplanets \citep[e.g.,][]{snellen2013}.  
Atomic oxygen is the third most abundant element in the 
universe (following H and He); however, it is still 
unknown what fraction of oxygen is 
contained within the deceptively simple \ce{O2} in interstellar and
circumstellar material.   

Gas-phase \ce{O2} has recently been observed in-situ in the coma of
comet 67P/Churyumov-Gerasimenko \citep[][hereinafter comet
67P/C-G]{bieler2015} by the ROSINA instrument on board the 
{\it Rosetta} spacecraft \citep[Rosetta Orbiter Spectrometer for Ion and
Neutral Analysis,][]{balsiger2007}. 
\ce{O2} is strongly correlated with \ce{H2O} and is present at an
average level of $3.8\pm0.85$\% relative to \ce{H2O}, making it the
fourth most abundant molecule in the comet,  following \ce{H2O},
\ce{CO2}, and \ce{CO}.     
The authors argue that \ce{O2} does not originate from gas-phase
chemistry in the coma but from direct sublimation from or within the
comet surface.   
The strong correlation with \ce{H2O} suggests that the \ce{O2} 
is trapped within the bulk \ce{H2O} ice matrix of the comet, 
which provides constraints concerning the chemical origin of the \ce{O2} ice.  
Processing of the cometary surface by solar wind particles and UV radiation 
has been ruled out by the authors, because the penetration depth 
(a few $\mu$m to m) is not sufficient to process material throughout the bulk. 
This process has been postulated to be responsible for the \ce{O2}-rich, 
yet tenuous, atmospheres of several of the icy moons of Saturn and 
Jupiter \citep[e.g.,][]{hall1995,spencer1995,teolis2010}.
Upon each pass into the inner solar system, comet 67P/C-G loses several meters 
of surface ice; hence, the surface revealed today is likely pristine.  
A reanalysis of data from the Neutral Mass Spectrometer on board 
the {\em Giotto} probe which did a fly-by of comet 1P/Halley in 1986, 
confirmed the presence of \ce{O2} at a level similar to that seen in
67P/C-G \citep{rubin2015b}.   
This suggests that \ce{O2} is not only an 
abundant molecule in comets, but is also common to both Jupiter-family comets, 
such as 67P/C-G, and Oort Cloud comets, such as 1P/Halley, which 
have different dynamical behaviours and histories.    

The 67P/C-G observations strongly suggest that \ce{O2} was present within the 
ice mantle on dust grains in the presolar nebula prior to comet formation.  
This then raises the question whether \ce{O2} was abundant 
in icy dust mantles entering the protoplanetary disk of the young Sun, 
or whether the conditions in the comet-forming zone of the early solar system
were favourable for \ce{O2} formation and survival.  
Upper limits on the abundance of \ce{O2} ice in molecular clouds
obtained with the {\it Infrared Space Observatory} (ISO) and
ground-based instruments are rather conservative
\citep[\ce{O2}/\ce{H2O}~$<0.6$,][]{vandenbussche1999, pontoppidan2003}. 
{\ce{O2} is a diatomic homonuclear molecule with zero electric dipole
moment; hence it does not possess  electric dipole-allowed rotational
transitions  which makes it difficult to detect in cold 
environments via remote sensing. 
Therefore, gas-phase \ce{O2} has been particularly elusive in interstellar
clouds, early attempts to detect gas-phase \ce{O2} in molecular clouds 
with the {\it Submillimeter Wave Astronomy Satellite} (SWAS) and {\it Odin} 
resulted in upper limits only, $\lesssim 10^{-7}$ relative to \ce{H2}
\citep{goldsmith2000,pagani2003}. }

More recent and higher sensitivity observations with {\em Herschel} 
allowed a deep search for \ce{O2} towards sources considered true
solar system progenitors: low-mass protostars.     
A deep upper limit was determined towards the well-studied 
protostar, NGC~1333-IRAS~4A, \citep[\ce{O2}/\ce{H2} $<
6\times10^{-9}$,][]{yildiz2013}.   
Detailed modelling of the chemistry throughout the well-characterised envelope 
of IRAS~4A demonstrates that the material entering the 
protoplanetary disk, both gas and ice, is likely poor in molecular
oxygen.  
For a \ce{H2O}/\ce{H2} abundance of $\sim 5 \times 10^{-5}$, the
inferred limit would correspond to a \ce{O2}/\ce{H2O} abundance ratio
of $\leq 0.012$ \%. 
This picture is consistent with laboratory experiments that 
have shown that \ce{O2} ice is efficiently hydrogenated at 
low temperatures and converted into \ce{H2O} and \ce{H2O2} ices
\citep[$\lesssim 30$~K,][]{ioppolo2008,miyauchi2008}.  
This makes the close association of \ce{O2} with \ce{H2O} in 67P/C-G 
an even stronger enigma.  

However, {\em Herschel} did reveal the presence of gas-phase \ce{O2}
in two sources:  
Orion \citep[$\ce{O2}/\ce{H2}\approx
0.3-7.3\times10^{-6}$,][]{goldsmith2011, chen2014} 
and $\rho$~Oph~A \citep[$\ce{O2}/\ce{H2}
\approx5\times10^{-8}$,][]{larsson2007,liseau2012}.   
Orion is a region of active star formation and the location of the 
gas-phase \ce{O2} emission coincides with a clump of very warm ($65 -
120$~K) and dense gas, a so-called \ce{H2} `hot spot', which may have
recently been subjected to shocks \citep[e.g.,][]{melnick2015}.  
These conditions are not representative of those expected 
in the molecular cloud from which the Sun formed.  
On the other hand, $\rho$~Oph~A is a dense core in the more quiescent 
$\rho$~Oph molecular cloud complex, which stands out  
from other low-mass star-forming regions by exhibiting emission from relatively 
warm molecular gas \citep[$\gtrsim 20$~K,][]{liseau2010,bergman2011a}. 
Subsequent observations of $\rho$~Oph~A have also determined the presence 
of related gas-phase species, \ce{HO2} and \ce{H2O2}, at an abundance 
level on the order of $2\times10^{-3}$ that of \ce{O2} \citep{bergman2011b,parise2012}. 
These molecular ratios show reasonable agreement with those seen 
in 67P/C-G with ROSINA 
\citep[$\ce{HO2}/\ce{O2} = (1.9\pm0.3) \times 10^{-3} $,
$\ce{H2O2}/\ce{O2} = (0.6\pm0.07) \times 10^{-3}$,][]{bieler2015}.   
The chemically related species, \ce{O3} (ozone), was not detected in the comet coma 
with a very low upper limit, $< 2.5\times 10^{-5}$ with respect to \ce{O2}.  

In summary, despite \ce{O2} being a particularly elusive molecule in interstellar 
and circumstellar environments, there apparently do exist conditions which 
are favourable for the formation of \ce{O2} and related species 
at abundance ratios similar to that observed in ices in comet 67P/C-G.  
{By assuming that all the energy deposited into water ice by high
  energy particles is used to convert \ce{H2O} into \ce{O2},
\citet{mousis2016} claimed
that radiolysis of water-containing interstellar ices in molecular
clouds is the only mechanism that produces \ce{O2} in high abundances.
However, laboratory experiments of cold interstellar ice analogs 
show that \ce{O2} can also be efficiently formed through non-energetic
surface chemistry before being converted to water
\citep[see][]{minissale2014a} while the production of \ce{O2} through
water radiolysis should be accompanied by a more efficient production
of \ce{H2O2}, in contradiction with the low abundance of \ce{H2O2}
observed in 67P/C-G.  

Here we investigate the formation and survival of \ce{O2} ice
using a variety of sophisticated astrochemical models, taking an
extended chemical network including the formation and destruction
pathways of \ce{O2} into account, in order to elucidate the
origin of  cometary \ce{O2}, and help explain its strong correlation
with water ice and the low abundances of its chemically related species.}    
We explore and discuss several different origins:  
i) \ce{O2} synthesis in ice mantles in dark clouds (``primordial'' origin), 
ii) \ce{O2} formation and survival {\it en route} from the protostellar envelope 
into the disk and subsequent delivery into the comet-forming zone, and 
(iii) {\it in-situ} formation of \ce{O2} within the protoplanetary disk prior to 
comet formation.  
This work differs from that presented in \citet{mousis2016} because we
consider all possible chemical  pathways between \ce{O2} and other
O-bearing species, including \ce{H2O}, \ce{HO2}, \ce{H2O2}, and \ce{O3}.
In Section~\ref{chemistry} we describe the interstellar chemistry of
molecular oxygen, in Sections 3 to 5 we systematically discuss each
scenario, presenting  the necessary evidence for or against each hypothesis, 
and in Section~\ref{summary} we summarise our main findings.

\section{Interstellar chemistry of {O$_2$}}
\label{chemistry}

\begin{figure*}
\centering
\includegraphics[width=100mm]{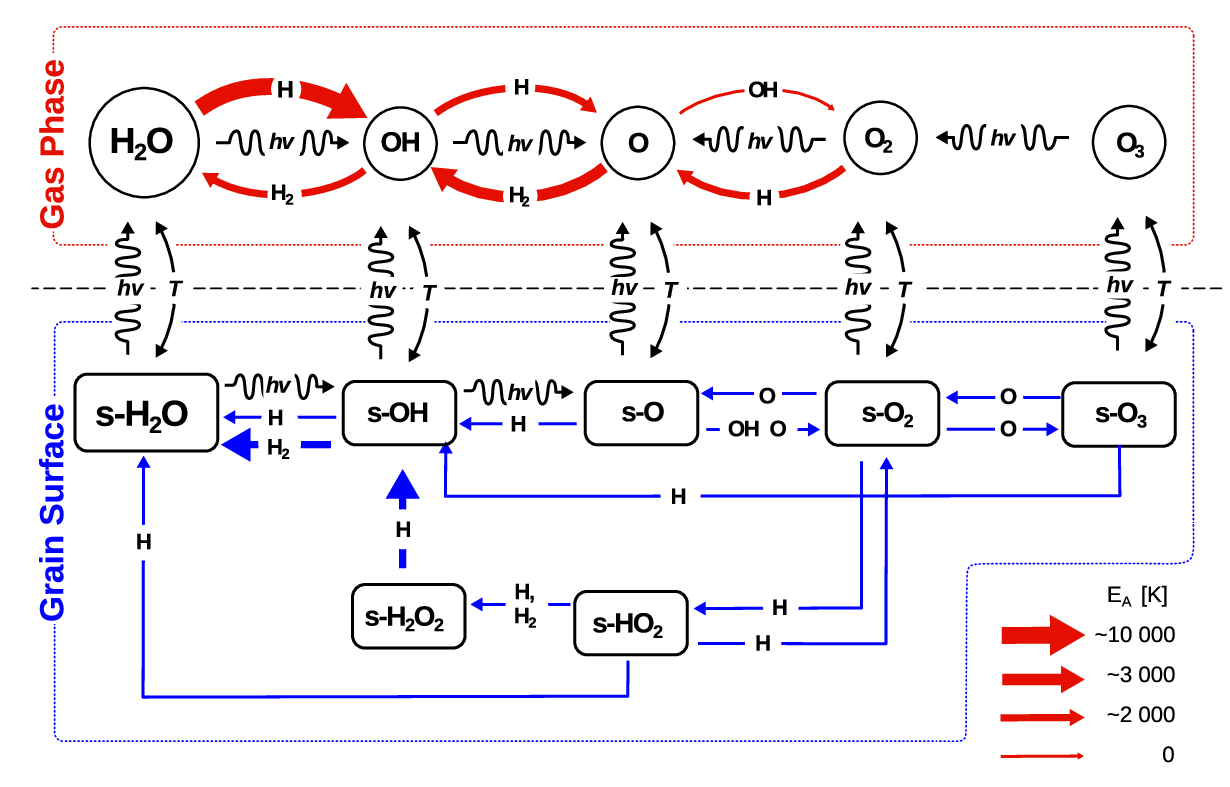}
\caption{
Summary of the main gas-phase and solid-state chemical reactions
leading to the formation and the destruction of molecular oxygen.
{ Gas phase neutral-neutral reactions have activation barriers whose
  values are estimated by the thickness of the arrow. s-X denote
  species X on the ice surfaces.}}
\label{scheme_O2}
\end{figure*}

Two main processes have been invoked for the formation of
molecular oxygen in the interstellar medium: i) gas-phase formation 
via neutral-neutral chemistry, and
ii) formation via association reactions on/within icy mantles of dust grains. 
The observations towards both $\rho$~Oph~A and 
67P/C-G, in conjunction with known chemical pathways studied in the 
laboratory, present several challenges for astrochemical models.  
First, the reproduction of the relatively high \ce{O2}/\ce{H2O} ice ratio 
simultaneously with the very low \ce{O3}/\ce{H2O} ice ratio, 
and second, the ratios of \ce{HO2}/\ce{O2} and \ce{H2O2}/\ce{O2} produced 
in the gas phase, assuming that chemistry on or within the ice mantle 
is responsible for the observed gas-phase ratios.  
Figure \ref{scheme_O2} summarises the chemical reactions involved in the 
formation and destruction of molecular oxygen which 
are discussed here.  

\subsection{Gas-phase chemistry}
Gaseous \ce{O2} is thought to form primarily via the barrierless
neutral-neutral reaction between O and OH in cold and warm gas.
Due to its importance, this reaction has been well studied 
both experimentally and theoretically.  
The rate coefficient has a negligible temperature dependence, 
with a recommended value (based on theoretical calculations and
experiments) between $2 \times 10^{-11}$ and $8 \times
10^{-11}$~cm$^3$~s$^{-1}$ at 10~K, and  
an experimentally-constrained value of $7 \times
10^{-11}$~cm$^3$~s$^{-1}$ at 140 K decreasing to $3 \times
10^{-11}$~cm$^3$~s$^{-1}$ at 300 K \citep[see][for a discussion on the
rate coefficient]{hincelin2011}.   
The formation of \ce{O2} in cold dark clouds is initiated by the high
initial abundance assumed for atomic oxygen, 
inducing an efficient ion-neutral chemistry that also forms OH. 
In warm environments ($T\gtrsim100$~K), e.g., the inner regions of 
protostellar envelopes or the inner, warm layers of protoplanetary disks, 
OH and O are mostly produced through warm neutral-neutral chemistry driven 
by the photodissociation of water sublimated from interstellar ices. 
The gas-phase formation of the chemically-related species, \ce{O3}, 
is inefficient under interstellar conditions, as it requires 
three-body association of \ce{O2}~+~\ce{O} \citep{atkinson2004}; 
thus, despite this reaction possessing a negligible reaction 
barrier, it only proceeds under the high-density conditions found in 
planetary atmospheres and in the inner midplanes of protoplanetary 
disks.   

\subsection{Ice chemistry} \label{icechem}

Solid \ce{O2} in dark clouds is involved in the surface
chemistry reaction network leading to the formation of water ice
\citep[][]{tielens1982,cuppen2010,vandishoeck2013}.  
\ce{O2} is formed through atomic O recombination on ices and 
efficiently reacts with either atomic O or atomic H to form \ce{O3} or
\ce{HO2}, respectively, eventually leading to the formation of
water. 
The hydrogenation of \ce{O3} also leads to the formation of \ce{O2},  
in addition to dominating the destruction of \ce{O3} ice.

Laboratory experiments of interstellar ice analogues studying water
formation suggest that the  O~+~O, O~+~\ce{O2}, and H~+~\ce{O2}
reactions, involved in the formation and destruction of \ce{O2}, 
all have small or negligible reaction barriers. 
\citet{miyauchi2008} and \citet{ioppolo2008} independently 
studied the efficiency of the \ce{O2}~+~H reaction, with both studies 
concluding that this reaction is effectively barrierless, contradicting
the earlier quantum calculations by \citet{melius1979} for the 
gas-phase reaction which predicted an activation barrier of 1200~K. 
The reactivity of the O~+~O and O~+~\ce{O2} reactions is still a
matter of debate, and is discussed in section \ref{imp_params}.
The reaction, $\ce{O3}~+~\ce{O}\longrightarrow\ce{O2}~+~\ce{O2}$, 
is considered unlikely to occur on grain surfaces under 
dark cloud conditions because of its relatively high activation 
energy barrier, 2000~K, as experimentally determined for the gas-phase reaction 
\citep{atkinson2004}.  

Dark clouds, the inner regions of protostellar envelopes, and 
the comet-forming regions of protoplanetary disks, 
are all well-shielded from external sources of UV radiation 
($\mathrm{A}_\mathrm{v} \gtrsim 10$~mag); 
however, water ice can be photodissociated by 
cosmic-ray-induced UV photons produced by the excitation of
molecular hydrogen by electrons generated by cosmic-ray ionisation 
of \ce{H2} \citep{prasad1983}. 
Water ice photodissociation has been extensively studied in the laboratory 
\citep{westley1995,oberg2009} and in molecular dynamics (MD) simulations
\citep{andersson2006,andersson2008,arasa2015}. 
The MD simulations show that water ice which is photodissociated generates 
OH and H photoproducts that move through the ice due to their excess energy.   
Each photodissociation event can lead to various chemical outcomes 
(e.g., direct desorption into the gas phase or recombination followed by desorption or trapping), 
the probabilities for which are dependent upon the depth into the 
ice mantle \citep[and fully tabulated in][]{arasa2015}. 
The detection of \ce{O2} following the 
UV irradiation of cold water ice also supports water ice photodissociation 
into O~+~\ce{H2} or O~+~H~+~H photoproducts \citep{oberg2009,heays2016}. 

{ Laboratory experiments show that
the bombardment of cold water ices with ionizing energetic particles can result
in the formation of \ce{O2} and other chemically related species from
the destruction of water \citep[see][]{matich1993, sieger1998, baragiola2002,
  zheng2006, loeffler2006, teolis2010, hand2011}.
The production of \ce{O2} and \ce{H2O2} through irradiation of water
ice by energetic particles depends on the projectile penetration
depth. Low-energy ions, for example, only penetrate the few dozen
outermost ice layers, where H and \ce{H2} can easily escape, favouring an
efficient production of \ce{O2} relative to \ce{H2O2}. The yield of
O$_2$ production therefore tends to decrease with the energy of the
irradiating particles from a few $10^{-3}$ molecule eV$^{-1}$ for
keV protons to $10^{-6}$ for MeV ions \citep[see][]{teolis2010}. 
 }
{ Irradiation of energetic ions during the condensation of water molecules
  can dramatically enhance the production of \ce{O2} up to
  \ce{O2}/\ce{H2O} abundances ratios of $\sim 30$ \%
  \citep{teolis2006}. 
}

\subsection{Gas-ice balance} \label{gasice}

O$_2$ formed in the ice mantle under dark 
cloud conditions ($T\sim10$~K) can be returned to the gas-phase 
via a multitude of non-thermal desorption processes \citep[e.g.,][]{tielens2013}.  
Those mechanisms which have been quantified in the laboratory for \ce{O2} 
include photodesorption by cosmic-ray-induced UV photons 
\citep{fayolle2013,zhen2014}, 
and desorption induced by exothermic chemical reactions 
\citep[i.e., chemical desorption,][]{minissale2014b,minissale2016}.  
Photodesorption of \ce{O2} was found to be triggered by 
photodissociation, with \ce{O2} returned to the gas-phase with 
yields of $\sim 10^{-3}$~molecules per incident photon, for a radiation 
spectrum appropriate for the cosmic-ray-induced UV field and pure
\ce{O2} ice \citep{fayolle2013}. 
\ce{O3} is also detected in the experiments by \citet{zhen2014} with 
yields a factor of a few lower than those for \ce{O2}.  
\ce{O3} is not seen in the experiments by \citet{fayolle2013} due to the 
lower FUV fluences in the synchrotron experiments.  

{The probability of chemical desorption depends strongly on the type of
reaction and on the substrate and can vary between 0  and 80\%.
The chemical desorption efficiency of the O~+~O reaction was found to be
$\approx80$\% in experiments of \ce{O2} formation via oxygen
recombination on bare olivine-type surfaces \citep{minissale2014b}.
However, in experiments with higher oxygen coverage, the efficiency was
reduced to an estimated upper limit of  $\approx 5$\% probably due to
an efficient dissipation of the energy released  by the exothermic
reaction into the water ice \citep{minissale2014b,minissale2016}. 
The standard chemical desorption efficiencies assumed in this work are
the theoretical values computed by \citet{minissale2016} for the submonolayer
regime on bare grains. However, they should be regarded as upper
limits.
The O~+~O reaction has a high theoretical probability of 68\%
while reactions \ce{O2}~+~H, \ce{HO2}~+~H, and \ce{H2O2}~+~H show much lower
theoretical chemical desorption probabilities of 0.5 - 2 \%, in agreement with the
experimental upper limits. We explore in Section \ref{rhoopha}
the impact of the chemical desorption efficiencies on the gas phase
abundances of \ce{O2} and its chemically related species.
When data are not available, the chemical desorption probability is
fixed to 1.2\% \citep{garrod2007}.}

The binding energies of \ce{O2} to a variety of surfaces, including 
dust-grain analogues and water ice, have been measured in the
laboratory
\citep[$\approx900$~K,][]{collings2004,fuchs2006,acharyya2007,noble2012,collings2015}.   
This low binding energy makes \ce{O2} a particularly volatile 
species, expected to desorb at temperatures similar to CO.   
In temperature-programmed desorption (TPD) experiments with \ce{O2} 
layered on top of, and fully mixed with, water ice, 
a fraction of \ce{O2} is found to remain trapped within the ice matrix 
and released at higher temperatures \citep{collings2004}.  
The trapped fraction depends upon the deposition temperature 
with a greater fraction of volatiles trapped within the 
water ice when deposited at lower temperatures \citep[][]{collings2003}.  

\subsection{Important parameters for the chemistry of {O$_2$}} \label{imp_params}

The \ce{O2} formation and survival in dark clouds and protoplanetary disks
depends on a number of parameters, which are linked in turn to various
physical and chemical conditions: \\

1) The gas phase abundance ratio between H and O atoms that accrete onto
grains governs the competition between hydrogenation reactions leading
to \ce{H2O2} and \ce{H2O} and association reactions between O atoms,
forming \ce{O2} or \ce{O3} \citep{tielens1982}. 
For dark cloud conditions, the atomic H abundance in the gas phase is
a balance between its formation, which occurs via \ce{H2} ionisation
followed by dissociative electron recombination, and its conversion
back into \ce{H2} via recombination reactions on grain surfaces.   
At steady state and assuming a sticking probability of 1, the density
of H is therefore given by the ratio between these two processes
\citep{tielens2005} 
\begin{equation} 
  n(\textrm{H}) = \frac{2.3 \zeta n(\textrm{H}_2)}
{2 v(\textrm{H}) \sigma_d X_d n(\textrm{H}_2)} \label{eq_nH}
\end{equation}
where $v(\textrm{H})$ is the thermal velocity of atomic hydrogen,
$\zeta$ the cosmic ray ionisation rate, and $X_d$ and $\sigma_d$ the
abundance and the cross section of interstellar grains.  
The absolute number density of atomic H is therefore independent of the 
total density and increases linearly with the cosmic ray ionisation
rate.
Since the initial number density of atomic O increases linearly with the
total number density for a fixed oxygen abundance, the atomic H/O
abundance ratio increases (decreases) linearly with the total density
(cosmic-ray ionisation rate). 

2) The surface mobility of O atoms governs the reactivity of the O~+~O and
O~+~\ce{O2} reactions.  
The surface mobility of O atoms occurs mostly through thermal
hopping and depends exponentially on the dust temperature $T$, and their
diffusion energy $E_d$.
Astrochemical models which treat grain-surface chemistry 
usually scale the diffusion energy to the binding energy of the considered species $E_b(i)$, 
using a fixed value for the diffusion-to-binding energy ratio $E_d/E_b$ 
\citep[e.g.,][]{tielens1987}. 
As discussed by several authors \citep{cuppen2007,taquet2012}, 
$E_b$ and $E_d/E_b$ strongly depend upon the ice morphology and composition.      
The mobility of atomic oxygen on interstellar ice analogues has recently
been investigated by several experimental groups 
\citep{bergeron2008,he2015} who conclude that atomic O has a 
higher binding energy than the value of 800~K 
estimated by \citet{tielens1987}.
Theoretical calculations and experiments studying the
diffusion of molecules (CO or \ce{CO2}) or heavy atoms (O) on
several types of substrates suggest that species diffuse with low
diffusion-to-binding energy ratios of the order of $30-50$\% 
\citep{jaycock1986,karssemeijer2014}. 
However, experiments focusing on \ce{H2} formation via H
recombination on surfaces suggest a higher diffusion-to-binding 
energy ratio between 50 and 80\% \citep{katz1999,perets2005,matar2008}. 
The diffusion-to-binding energy ratio likely has a distribution of
values that depend upon the substrate (bare or ice-coated), 
the species under consideration (light atom, heavy atom, molecule), 
the ice morphology (porous, compact, crystalline, or amorphous ice), 
and the dominant composition of the chemically-active surface layer 
(\ce{H2O}, \ce{CO2}, or \ce{CO }).

3) The activation barriers of the O~+~O and O~+~\ce{O2} reactions 
directly govern the reactivity of the two reactions. 
\citet{minissale2014a} derive an upper limit of 150~K for the
reaction barrier for O~+~O and O~+~\ce{O2} in an
experimental study on an amorphous silicate surface. 
However, the presence of an activation barrier for the latter reaction
has been invoked by several authors \citep[see][]{dulieu2011}. 
For example, \citet{lamberts2013} require an activation
barrier of 500 K for the O~+~\ce{O2} reaction in order to reproduce the
results of laboratory experiments in
thick ices with their microscopic Monte-Carlo model. Here we explore
the effects of the parameter choices for these three key aspects of the
\ce{O2} chemistry.

\subsection{Astrochemical models} \label{astrochem}

Three state-of-the-art gas-grain astrochemical models have been used
in this work to study the formation and survival of molecular oxygen
from dark clouds to the Solar System: 
1) the multi-phase model by \citet{taquet2014} to study the formation
of \ce{O2} in dark clouds;
2) the multi-phase model by \citet{furuya2015} to study the formation
of \ce{O2} during the formation of protoplanetary disks;
3) the two-phase model by \citet{walsh2015} to study the formation of
\ce{O2} {\it in-situ} in protoplanetary disks.

The multi-phase gas-grain Taquet and Furuya models couple the
gas phase and ice chemistries with the approach developed by
\citet{hasegawa1993} to follow the multi-layer formation of
interstellar ices and to determine the gas-ice balance.  
Several sets of differential equations governing the time-evolution of 
abundances are considered: 
one for gas-phase species, one for surface ice-mantle species,
and one (or several) for bulk ice-mantle species. 
The equations governing chemical abundances on the ice surface and in the 
bulk ice are linked by an additional term that is proportional to the 
rate of growth or loss of the grain mantle. 
As a consequence, surface species are continuously trapped in the bulk 
because of the accretion of new species in dark clouds. 
Following \citet{vasyunin2013}, the chemically-active surface is 
limited to the top four monolayers.  
The bulk ice mantle is considered to be chemically inert.  
The original three-phase model considered in the Taquet model assumes
that the inert bulk ice mantle  has a uniform molecular composition.
In order to accurately follow the ice evolution in warm conditions,
the Furuya model considers a depth-dependent molecular 
composition, through the division of the 
inert bulk ice mantle into five distinct phases \citep[for details,
see][and references therein]{furuya2016}. 

{ Radiolysis, i.e. the bombardment of (ionizing) energetic particles depositing energy
into the ice, and/or photolysis, i.e. the irradiation of ultraviolet
photons breaking bonds, can trigger chemistry within
the bulk mantle of cold interstellar ices. }
We have investigated the impact of the UV photolysis induced by secondary
UV-photons on the bulk ice chemistry and the formation and survival of
\ce{O2} by activating the bulk chemistry and assuming the same ice
parameters as for the surface chemistry (same diffusion and binding
energies, same chemical reactions). In our model, the formation of
\ce{O2} from \ce{H2O} photodissociation is a multi-step process, starting
from the production of oxygen atoms from water or OH photodissociation
followed by their recombination.
We find that \ce{O2} cannot be efficiently produced in the bulk
through ice photolysis as the photodissocation of the main ice
components not only produces oxygen atoms, that recombine together to
form \ce{O2}, but also hydrogen atoms that react with \ce{O2} to reform
water even if \ce{H2O} ice photodissociation would go directly to \ce{H2}
rather than H since there are other molecules like \ce{CH4}, \ce{NH3} or
\ce{CH3OH} that produce hydrogen atoms that are very mobile.
Overall, activating the bulk chemistry decreases the abundance
of highly reactive species like O atoms or radicals but does not
affect the main ice species. 

{ 
Laboratory experiments show that \ce{O2} can be efficiently formed through
radiolysis of ices without overproducing \ce{H2O2} only if the radiolysis
occurs as water is condensing onto a surface \citep[][see
  section \ref{icechem}]{teolis2006}. 
However, in molecular clouds  water ice is mostly formed {\it in-situ} at the
surface of interstellar grains through surface reactions involving
hydrogen and oxygen atoms.  This happens prior to the formation of the
presolar nebula, i.e. the cloud out of which our solar system was
formed, and it is possible that the comet-forming zone of the 
Sun's protoplanetary disk inherited much of its water ice from the
interstellar phase \citep{visser2009, cleeves2014, altwegg2015, furuya2016}. 
 For the radiolysis mechanism to occur in the presolar
nebula, water ice would first need to be completely
sublimated and then recondensed prior to comet formation. 
  Luminosity outbursts induced by instabilities in the disk of
  the solar nebula can potentially provide a scenario for efficient
  \ce{O2} production in the ice matrix through sudden
  evaporation of water ice followed by fast recondensation. 
  We consider this scenario less likely because the cosmic-ray ionisation rate
  is thought to be impeded near the disk midplane with respect to
  interstellar values \citep[e.g.,][]{cleeves2013}.
Energetic ionizing particles from the (pre)solar wind are also expected to be
significantly attenuated close to the disk midplane by the intervening
large column of material ($>> 100$ g cm$^{-2}$) between the central star and
the comet-forming zone beyond $\sim 10$ AU. 

\citet{mousis2016} explored the \ce{O2} formation through radiolysis
of water within interstellar ices in the solar nebula to explain the high
abundance of \ce{O2} observed in comet 67P/C-G. However, they concluded that the
galactic cosmic-ray flux is not sufficient to produce the observed ratio of
\ce{O2}/\ce{H2O} over the lifetime of the presolar nebula.  }

The gas-phase chemical network used by the Taquet model is the 
non-deuterated version of that from \citet{taquet2014}, the basis for
which is the 2013 version of the KIDA chemical database
\citep{wakelam2012}. It has been further updated to include warm gas-phase
chemistry involving water and and ion-neutral reactions involving ozone. 
The network also includes the surface chemistry of all 
dominant ice components (\ce{H2O}, CO, \ce{CO2}, \ce{NH3}, \ce{CH4},
\ce{H2CO}, \ce{CH3OH}),  as well as those important for water (e.g.,
\ce{O2}, \ce{O3}, and \ce{H2O2}).   
Several new surface reactions were added involving \ce{O3} and reactive species 
such as N, O, OH, \ce{NH2}, and \ce{CH3}, following the NIST gas-phase
chemical database.

The gas-ice chemical network of \citet{garrod2006}, based on the OSU
2006 network, is used with the Furuya model.
 The gas phase and surface networks are more
suited to the high density and warm temperatures conditions found in
protostellar envelopes. It has therefore been supplemented with
high-temperature gas-phase reactions from \citet{harada2010} and includes the
formation of many complex organic molecules. It is consequently more
expansive than the network used in the Taquet model. 

{
The gas-phase chemical used in the Walsh model is based on the 2012
release of the UMIST Database for Astrochemistry
\citep[UDfA;][]{mcelroy2013}, supplemented by direct X-ray ionisation
reactions, X-ray-induced ionisation and dissociation processes, and
three-body reactions. The grain surface chemical network of
\citet{garrod2008} is used.}

Input parameters assumed for the three types of astrochemical models
are listed in Table \ref{input_table}. Unless otherwise stated, this
Table gives the standard values
for the physical parameters: the cosmic ray ionisation rate $\zeta$, the
flux of secondary UV photons; the grain surface parameters: the
dust-to-gass mass ratio $R_{\textrm{dg}}$, the grain diameter
$a_{\textrm{d}}$, the volumic mass of grains
$\rho_{\textrm{d}}$, the surface density $N_{\textrm{s}}$, the
diffusion-to-binding energy ratio $E_{\textrm{d}}/E_{\textrm{b}}$, the
number of chemically active monolayers $N_{\textrm{act}}$, and the
sticking coefficient of species heavier than H and \ce{H2}.
The elemental abundances of species correspond to the set EA1 from
\citet{wakelam2008}.

\begin{table}
\centering
\caption{Input parameters assumed in all astrochemical simulations.} 
\begin{tabular}{l c}
\hline
\hline
Input parameters		&	Values	\\
\hline				
\multicolumn{2}{l}{Standard physical parameters}				\\
\hline
$\zeta$	(s$^{-1}$)	&	$10^{-17}$	\\
$F$(sec. UV)	(cm$^{-2}$ s$^{-1}$)	&	$10^4$	\\
\hline				
\multicolumn{2}{l}{Grain surface parameters}				\\
\hline				
$R_{\textrm{dg}}$		&	0.01	\\
$a_{\textrm{d}}$	($\mu$m)	&	0.2	\\
$\rho_{\textrm{d}}$	(g cm$^{-3}$)	&	3	\\
$N_{\textrm{s}}$	(cm$^{-2}$)	&	$10^{15}$	\\
$E_{\textrm{d}}/E_{\textrm{b}}$		&	0.5	\\
$N_{\textrm{act}}$	(MLs)	&	4	\\
$S$ (heavy species)		&	1	\\
\hline				
\multicolumn{2}{l}{Initial abundances}				\\
\hline				
$X$(\ce{H2})		&	0.5	\\
$X$(\ce{He})		&	0.09	\\
$X$(\ce{C})		&	$7.30 \times 10^{-5}$	\\
$X$(\ce{N})		&	$2.14 \times 10^{-5}$	\\
$X$(\ce{O})		&	$1.76 \times 10^{-4}$	\\
$X$(\ce{Si})		&	$8.0 \times 10^{-9}$	\\
$X$(\ce{S})		&	$8.0 \times 10^{-8}$	\\
$X$(\ce{Fe})		&	$3.0 \times 10^{-9}$	\\
$X$(\ce{Na})		&	$2.0 \times 10^{-9}$	\\
$X$(\ce{Mg})		&	$7.0 \times 10^{-9}$	\\
$X$(\ce{Cl})		&	$1.0 \times 10^{-9}$	\\
\hline				
\end{tabular}
\label{input_table}
\end{table}

\section{Dark cloud origin?}

Here we investigate whether the \ce{O2} observed in 67P/C-G
has a dark cloud origin, using the chemistry of \ce{O2} ice and gas
described in the previous Section.     
For this purpose, we use the Taquet astrochemical model presented in
section 2.4. 
The Appendix presents a first parameter study, in which 
several surface and chemical parameters are varied, in order to 
reproduce the low abundances of the chemically related species
\ce{O3}, \ce{HO2}, and \ce{H2O2} with respect to \ce{O2} seen in comet
67P/C-G. The low abundance of \ce{O3} and \ce{HO2} relative to
\ce{O2} can be explained when a small activation barrier of $\sim$ 300
K is introduced for the reactions O~+~\ce{O2} and H~+~\ce{O2}, in
agreement with the Monte-Carlo modelling of
\citet{lamberts2013}. However, the abundance of \ce{H2O2} is still
overproduced by one order of magnitude, suggesting that other chemical
processes might be at work.  
A second parameter-space study is then conducted to determine the range of physical conditions 
(e.g., dust temperature, number density, and cosmic-ray ionisation rate) over which \ce{O2} ice and gas 
(and those for chemically-related species, \ce{O3}, \ce{HO2}, and \ce{H2O2}) 
reach abundances (relative to water ice) similar to that seen in 67P/C-G.  
Finally, the case of $\rho$~Oph~A, where gas-phase \ce{O2} has been detected 
in the gas phase, is revisited with the same chemical model.

\subsection{Impact of physical and chemical parameters} \label{modelgrid}

The low temperature, in conjunction with the low flux of UV photons 
found in interstellar dark clouds, promotes the formation of interstellar ices.
The ice chemical composition depends on various physical and chemical
parameters as discussed in section \ref{imp_params}. 
To investigate the formation and survival of \ce{O2} under dark cloud 
conditions, a model grid is run in which the total density of H
nuclei, $n_{\textrm{H}}$, the gas and dust temperature, $T$ (assumed
to be equal), the cosmic ray ionisation rate, $\zeta$, and the visual
extinction, $A_{\textrm{V}}$ are varied following the methodology 
described in \citet{taquet2012}.  
Values explored in the model grid are listed in Table
\ref{grid_table}, resulting in 500 models in total. 
In these models, the ``standard'' set of chemical parameters derived
in the Appendix are assumed (see Table \ref{grid_table}).  


\begin{figure}
\centering
\includegraphics[width=\columnwidth]{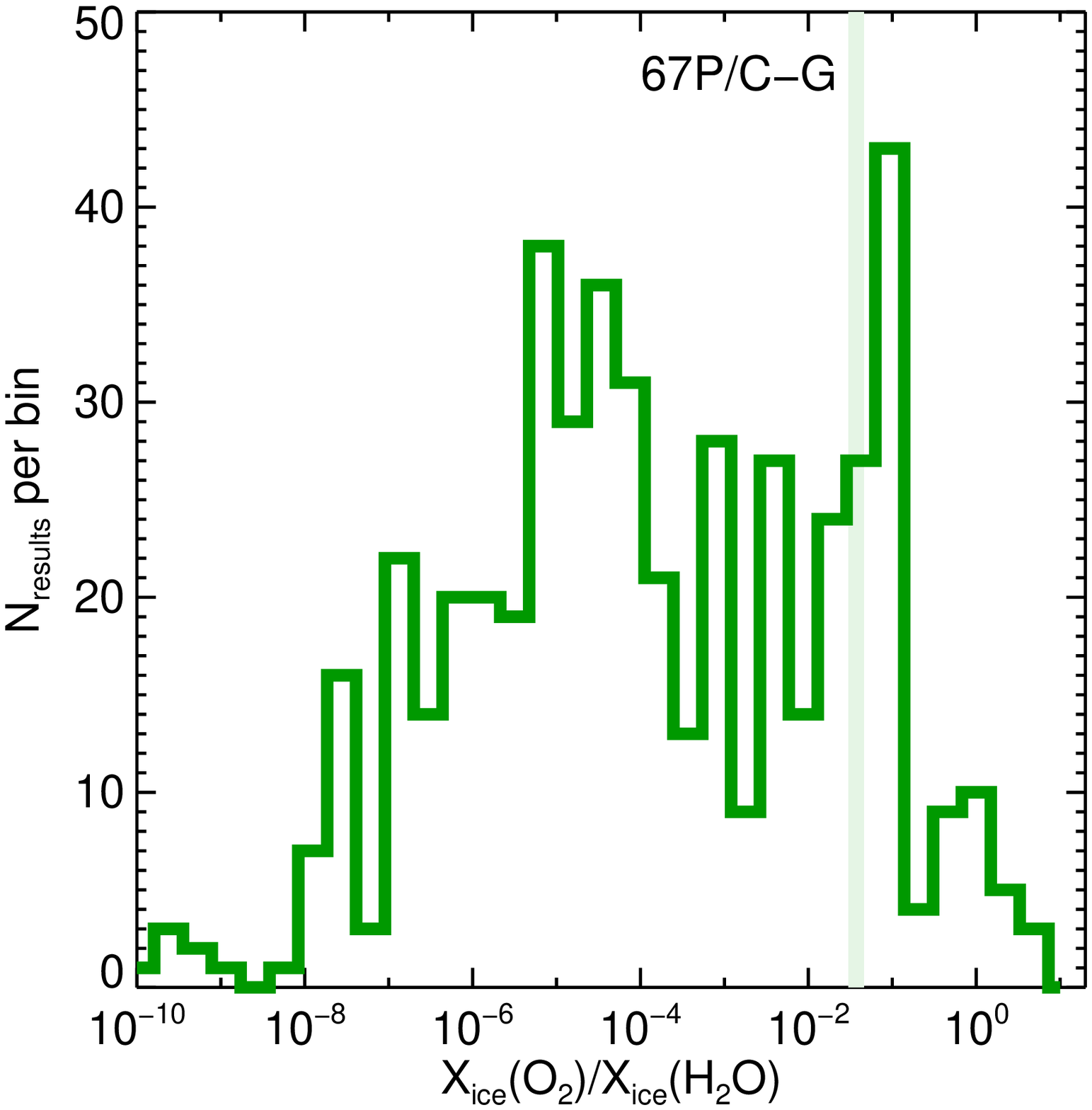}
\includegraphics[width=\columnwidth]{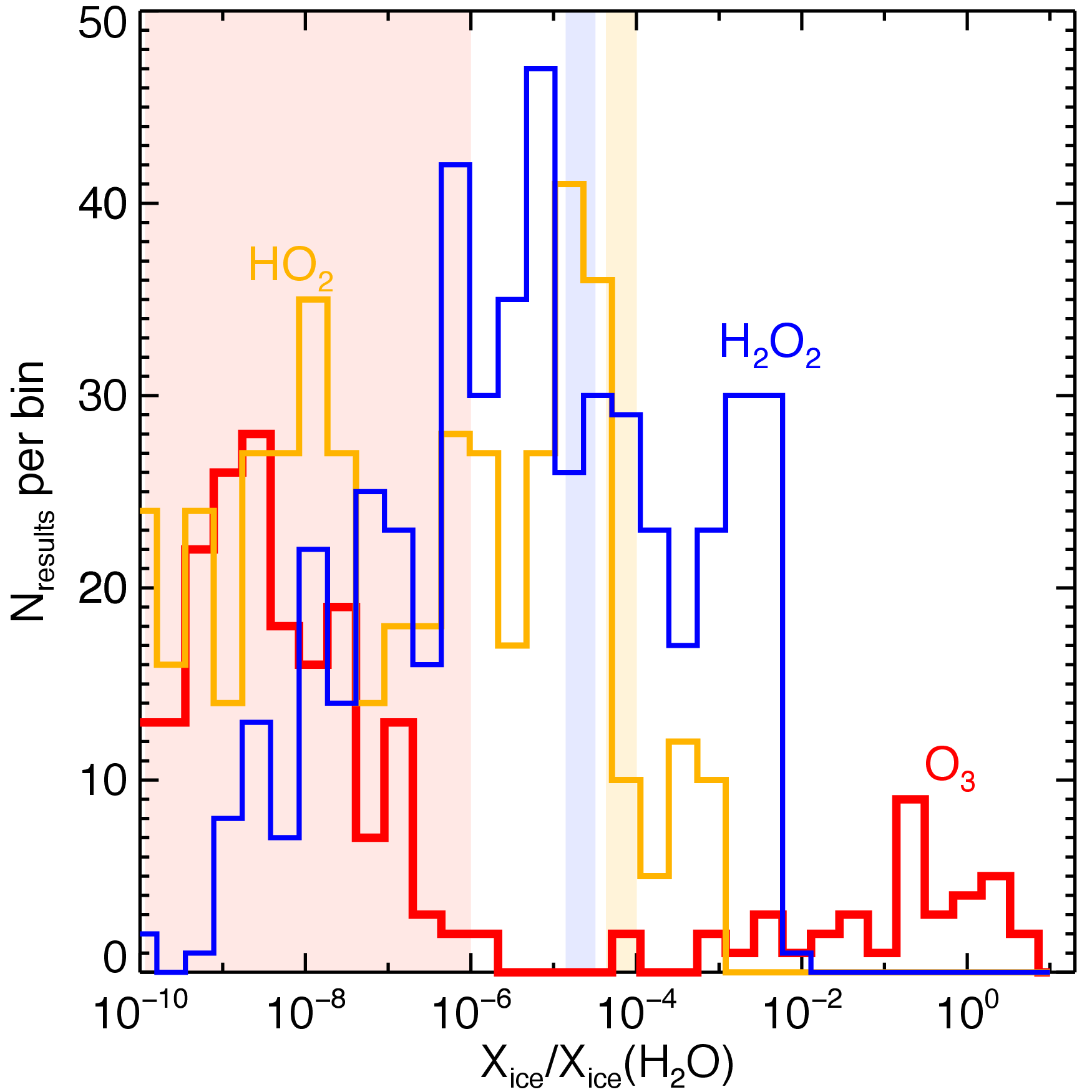}
\caption{
Distribution of final abundances of solid \ce{O2} (green, top panel),
and \ce{O3} (red), \ce{HO2} (yellow), and \ce{H2O2} (blue, bottom
panel) relative to water ice at the free-fall time  (defined in the
text), for the complete model grid { in which the total density, the
  temperature, the cosmic ray ionisation rate, and the visual
  extinction are varied within the range of values given in Table
  \ref{grid_table}} (see Section~\ref{modelgrid}). 
The thick dashed lines or the solid boxes refer to the abundances
observed in the comet 67P/C-G.}
\label{modelgrids1}
\end{figure}

The abundances of all species in the reaction network 
are evolved from their assumed initial abundances 
(see Section~\ref{astrochem}) as a function of time only, i.e., 
assuming constant physical conditions.  
Figure \ref{modelgrids1} shows the distribution of abundances of solid
\ce{O2}, and the chemically related species, \ce{O3}, \ce{HO2}, and \ce{H2O2}, 
relative to water ice, at the free-fall time, $t_\mathrm{FF}$, defined as 
\begin{equation}
t_\mathrm{FF} = \sqrt{\frac{3\pi}{32 G n_\ce{H} m_p}} \quad \mathrm{s}, 
\end{equation} 
where $G$ is the gravitational constant and $m_p$ is the proton mass.  
$t_\mathrm{FF}$ varies across the grid from $4.4 \times 10^4$ to $1.4
\times 10^6$ yr. 
Cores can have longer lifetimes, e. g. due to magnetic support, up to
10 $t_\mathrm{FF}$. However, assuming a longer timescale does not
change our conclusions because interstellar ices form in a
timescale similar to $t_\mathrm{FF}$. 
The results show that the formation and survival of solid \ce{O2}, and
other reactive species, in interstellar ices, is  
strongly dependent upon the assumed physical conditions.  
The model grid shows a large dispersion of final abundances of solid \ce{O2}
from $< 10^{-10}$ to $10$ relative to water ice 
(top panel of Figure~\ref{modelgrids1}). 
Due to its lower reactivity, hydrogen peroxide, \ce{H2O2}, shows 
a slightly more narrow final abundance dispersion, with most of the
models predicting values between $10^{-6}$ and $10^{-2}$ (1 \%) with
respect to water ice (see bottom panel of Fig. \ref{modelgrids1}). 
\ce{HO2} is mostly formed in the ice mantle via the hydrogenation of \ce{O2}, 
and is converted into \ce{H2O2} via a subsequent barrierless
hydrogenation reaction, \ce{O2} being a precursor of \ce{H2O2};  
hence, its final abundance is governed by that of \ce{O2} ice, and
therefore follows a similar trend but lower by four orders of
magnitude due its high reactivity. 
Ozone, formed from molecular oxygen via the \ce{O2}~+~O reaction also
displays a broad distribution of abundances but most of the models
predict abundances lower than $10^{-6}$ relative to water, due to the
small O~+~\ce{O2} barrier. 

\begin{figure*}
\centering
\includegraphics[width=140mm]{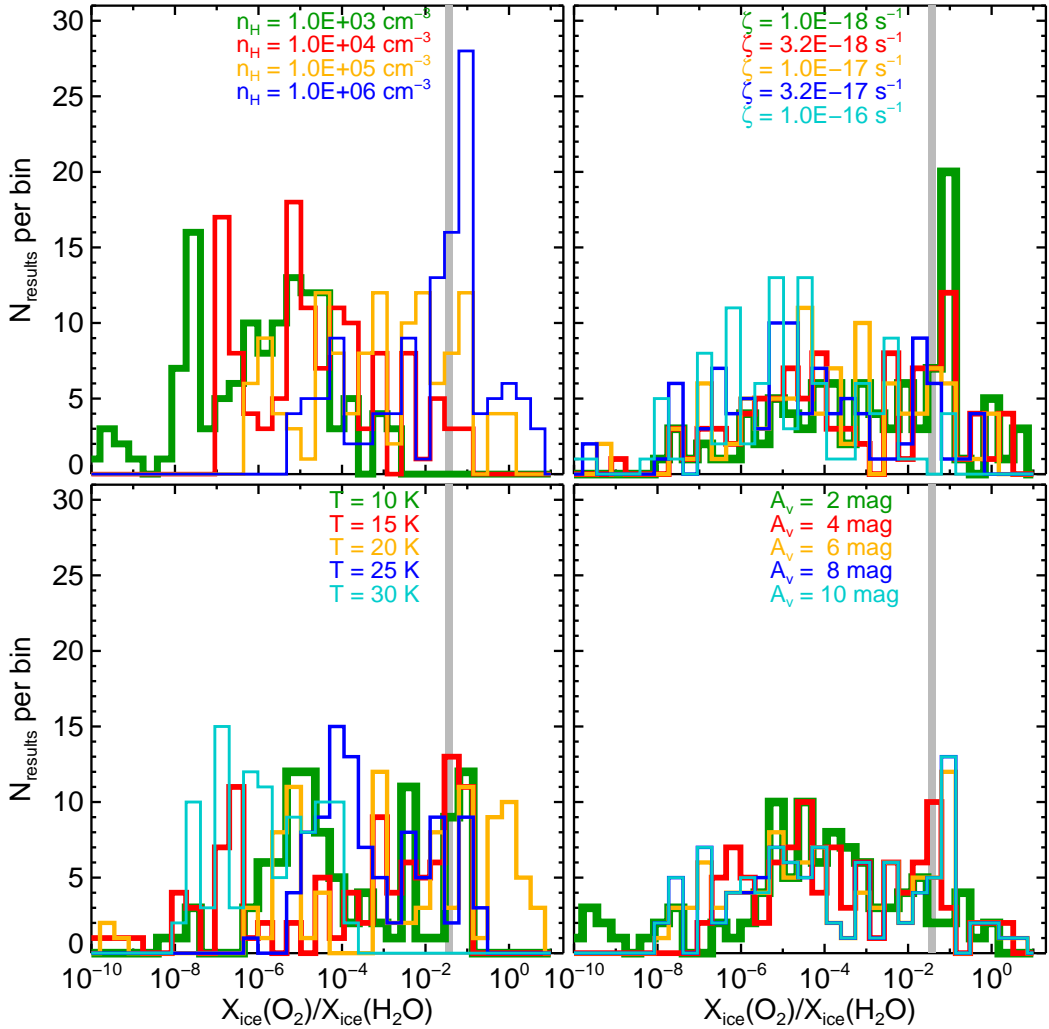}
\caption{
Distribution of final abundances of solid \ce{O2} relative to water ice 
at the free-fall time (defined in the text), for the range of densities (top left), 
temperatures (bottom left), cosmic-ray ionisation
rates (top right), and visual extinctions (bottom right), assumed in
the model grid (see Section~\ref{modelgrid}). { For each panel, the
``standard'' values of other parameters, listed in Table
\ref{grid_table}, are assumed.}
The grey solid boxes refer to the
\ce{O2} abundance observed in the comet 67P/C-G.}
\label{modelgrids2}
\end{figure*}

Figure \ref{modelgrids2} shows the distribution of the 
final abundance of solid \ce{O2} relative to water ice,  
for the range of assumed values for each
physical parameter varied in the model grid. 
High \ce{O2} abundances ($>4$\% relative to water ice) are obtained only for those 
models with high densities ($n_{\textrm{H}} \gtrsim 10^5$ cm$^{-3}$).  
As discussed in Section \ref{imp_params}, higher gas densities result
in a lower gas-phase H/O ratio,  thereby increasing the rate of the
association reaction between  O atoms to form \ce{O2} ice, and
correspondingly decreasing the rate  of the hydrogenation reactions,
O~+~H and \ce{O2}~+~H,  which compete with \ce{O2} ice formation, and
destroy \ce{O2} ice once formed,  respectively.    

An intermediate temperature of 20~K is also favoured because it
enhances the mobility of oxygen atoms on the grain surfaces whilst at the same time 
allowing efficient sublimation of atomic H.   
This additionally enhances the rate of oxygen recombination 
forming \ce{O2}, with respect to the competing hydrogenation
reactions.
Models with lower temperatures of 10 or 15 K can also reproduce the
\ce{O2}/\ce{H2O} of 4\% if a high density of $n_{\rm H} \sim 10^6$
cm$^{-3}$ is considered. 
Moreover, because the density of gas-phase H atoms increases linearly with the
cosmic-ray ionisation rate, $\zeta$, a low value of $\zeta$ also tends to
favour the survival of \ce{O2} ice. 
On the other hand, the visual extinction does not have a
strong impact on the abundance of solid \ce{O2} as the distributions of
abundances obtained for the five visual extinction values are very
similar. Thus, the final \ce{O2} ice abundances depend more 
strongly upon the assumed gas density, temperature, and
cosmic-ray ionisation rate, and high \ce{O2} ice abundances occur when
the initial atomic H/O ratio is low ($\leq 10^{-2}$). 

\begin{figure}
\centering 
\includegraphics[width=\columnwidth]{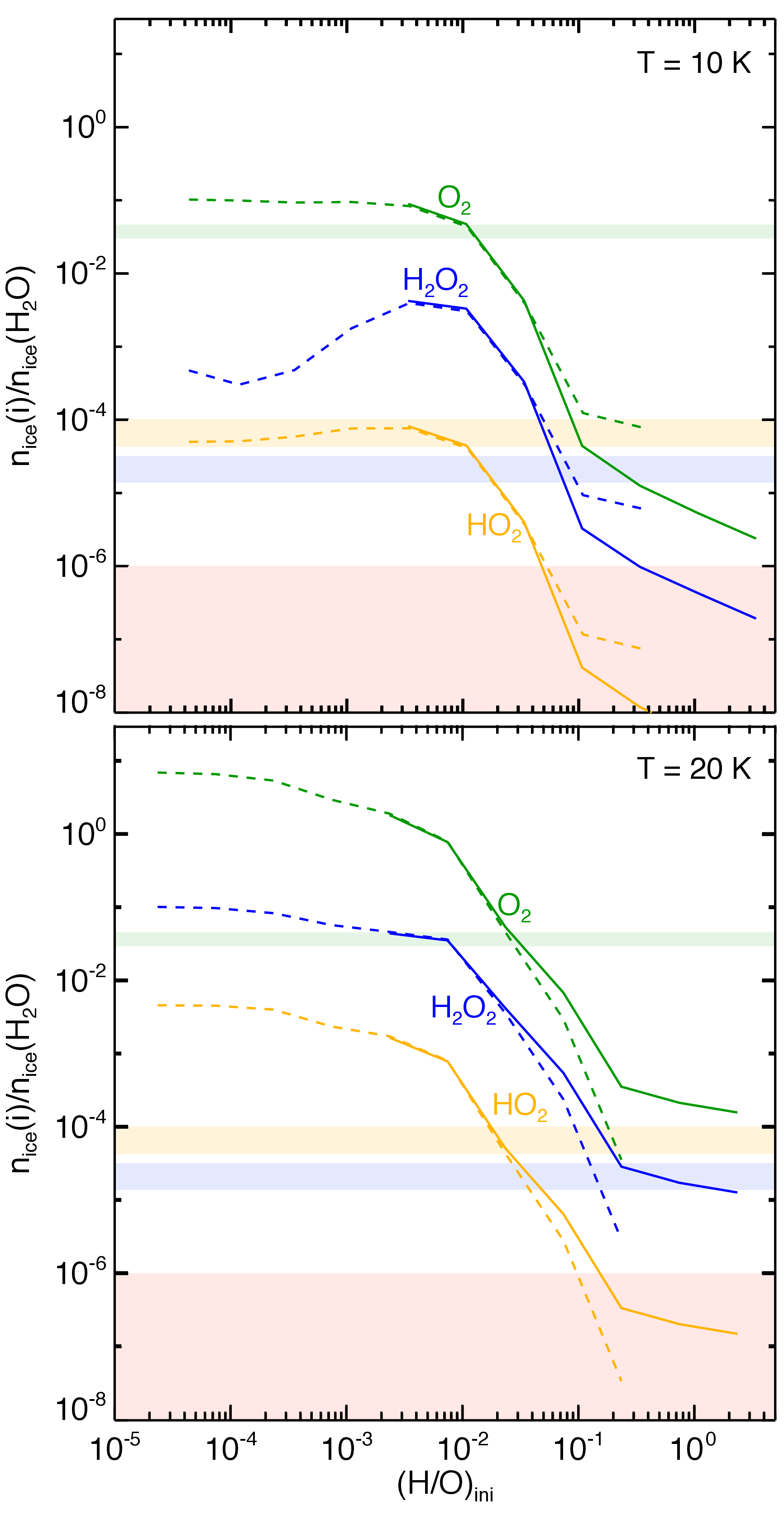}
\caption{Final abundances of \ce{O2}, \ce{O3}, \ce{HO2}, and \ce{H2O2} in
  interstellar ices with respect to water as function of the initial
  H/O abundance ratios given by different cosmic ray ionisation rates (and assuming 
  $n_{\textrm H} = 10^6$ cm$^{-3}$, dashed lines) and different
  densities (and assuming $\zeta = 10^{-17}$ s$^{-1}$, solid lines) at
  $T = 10$ K (top) and $T = 20$ K (bottom). 
{ The ``standard'' values of other parameters, listed in Table
\ref{grid_table}, are assumed.}
The solid boxes refer to the abundances
observed in comet 67P/C-G.} 
\label{HOratio}
\end{figure}

To illustrate further the crucial impact of the density and the cosmic
ray ionisation rate on the chemical composition of ices, 
Figure \ref{HOratio} shows the evolution of the abundances of \ce{O2}
and its chemically related species with respect to water ice as a function
of the initial atomic H/O abundance ratio induced by a variation of
the total density (assuming a constant $\zeta$ of $10^{-17}$ s$^{-1}$) or a
variation of the cosmic ray ionisation rate (assuming $n_{\textrm H} =
10^6$ cm$^{-3}$) at $T = 10$ and $20$ K. 
According to equation (\ref{eq_nH}), the initial atomic H/O abundance ratio
follows the expression 
\begin{equation}
\left( \dfrac{\ce{H}}{\ce{O}} \right)_{\rm ini} = 3.4 \times 10^{-3} 
\frac{\zeta}{10^{-17}~{\rm s}^{-1}}
\frac{10^6~{\rm cm}^{-3}}{n_{\ce{H}}}
\frac{1.76 \times 10^{-4}}{X({\ce{O}}_{\rm ini})}
\sqrt{\frac{10~{\rm K}}{T}}
\label{HOratio_eq}
\end{equation}
assuming the grain parameter values listed in Table
\ref{input_table}. 
For each temperature case, the evolution of the abundance ratios with
the initial atomic H/O abundance ratio follows similar trends, suggesting that the
initial atomic H/O abundance ratio, and consequently the $n_{\rm
  H}$/$\zeta$ ratio, is the dominant parameter for the
formation and survival of \ce{O2} and its chemically related species
in dark clouds.
The formation of \ce{O2} ice is strongly inhibited ($\ce{O2}/\ce{H2O}
\lesssim 1$\%) for high initial H abundances ([H]/[O]$_{\textrm{ini}}
\gtrsim 5 \times 10^{-2}$) induced by high cosmic-ray ionisation rates
and/or low densities, as it increases the rate of conversion of \ce{O2} ice to \ce{H2O}
ice. 
{For low cosmic-ray ionisation rates or high densities inducing initial
H/O ratios lower than $10^{-2}$, the formation of H atoms in the gas
phase is no longer dominated by \ce{H2} ionisation followed by
dissociative recombination but by neutral-neutral reactions involving O
atoms. The abundances of \ce{O2} and other chemically-related species
are consequently no longer influenced by $\zeta$ nor $n_{\rm H}$ and
remain constant. }
The results here demonstrate that a high abundance of \ce{O2}, 
at a level similar to that measured in 67P/C-G, seems to require an
initial H/O abundance ratio lower than $\sim 2-3 \times 10^{-2}$
(depending on the temperature) or, according
to equation (\ref{HOratio_eq})
\begin{equation}
 \frac{n_{\ce{H}}}{\zeta} \geq 10^{22}~{\rm cm}^{-3}~{\rm s}
\end{equation}
assuming the initial abundances listed in Table \ref{input_table}.

\begin{figure}
\centering 
\includegraphics[width=\columnwidth]{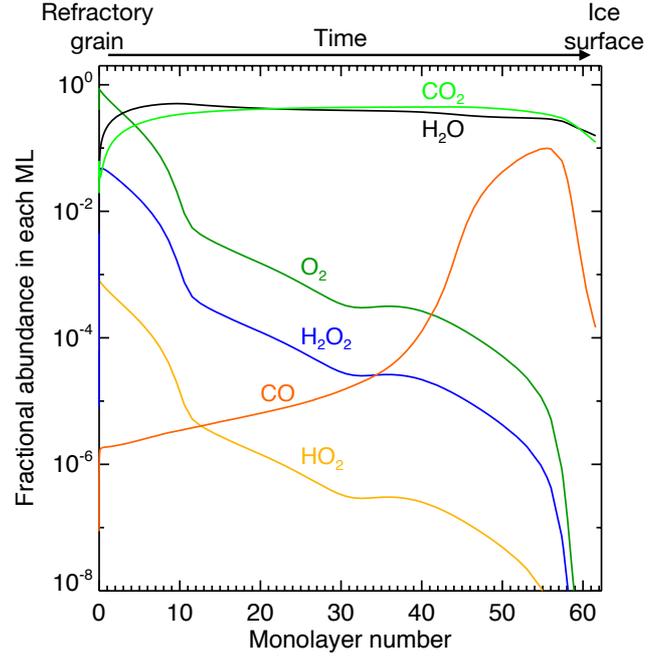}
\caption{Fractional composition of each ice monolayer as function of
  the monolayer number or ice thickness for the model that best
  reproduces the observations of comet 67P/C-G.
}
\label{poplay}
\end{figure}

Figure \ref{poplay} shows the chemical composition of the ice
obtained for the model using the physical conditions that best
reproduce the observations in comet 67P/C-G  ($n_{\textrm H} = 10^6$
cm$^{-3}$, $T = 21$ K, $\zeta = 10^{-16}$ s$^{-1}$), and the chemical
parameters derived in the Appendix. 
The fractional composition in each ice monolayer is plotted as
function of monolayer number,  i.e. the ice thickness that grows with time. 
At such a high density ($10^{6}$~cm$^{-3}$), hydrogenation reactions are less efficient 
due to the lower relative abundance of atomic H, and the freezeout 
timescales are sufficiently fast that reactive species can be 
trapped in the ice mantle before conversion into more stable molecules, 
like \ce{H2O}.  
The higher temperature (21~K) also enhances the mobility of heavier 
species, such \ce{O}, to increase the relative abundance of ice
species such as \ce{O2}  and \ce{CO2}.  
As a consequence, the most abundant species are water and carbon
dioxide. 
\ce{O2} ice is mostly present in the innermost layers of the ice mantle
and decreases in relative abundance towards the ice surface,
reflecting the initial low ratio of H/O in the gas phase, but tends to
be well mixed with \ce{H2O} ice.  In contrast, \ce{CO} is mostly formed in
the outer part of the ices, allowing an efficient sublimation,
explaining { its weak correlation with water in 67P/C-G. }

\subsection{The $\rho$ Oph A case} \label{rhoopha}

The $\rho$~Oph~A core, located at a distance of 120~pc, constitutes
the best test case for the water surface network and the production of
\ce{O2} in dark clouds because  it is the only interstellar source so far where
gas-phase \ce{O2}, \ce{HO2}, and \ce{H2O2} have been detected
\citep{bergman2011b, liseau2012, parise2012}.  
The parameter study presented in the previous Section suggests that
the physical conditions of $\rho$~Oph~A, a high density
($n_{\textrm{H}} \sim 10^6$ cm$^{-3}$), and  a relatively warm gas
temperature ($T_{\textrm{kin}} = 24 - 30$ K) and dust temperature
($T_{\textrm{dust}} \sim 20$ K), derived by \citet{bergman2011a} are
consistent with those which facilitate the formation and survival
of \ce{O2} ice. 

\ce{O2}, \ce{O3}, \ce{HO2}, and \ce{H2O2} are mostly, and potentially only, produced via
surface chemistry; hence their gas-phase abundances depend on their
formation efficiency  in interstellar ices and on the probability of
desorption upon formation  through chemical desorption (which is 
the dominant non-thermal  desorption mechanism for these species
in dark cloud conditions).   
As explained in Section \ref{gasice}, the chemical desorption
probabilities assumed in this work are the theoretical values computed
by \citet{minissale2016} and \citet{cazaux2016} for more than 20
reactions involved in the water and methanol chemical networks and
vary between 0 and 70\%. 
When data are not available, the chemical desorption probability is
fixed to 1.2\% \citep{garrod2007}. 

\begin{figure}
\centering 
\includegraphics[width=\columnwidth]{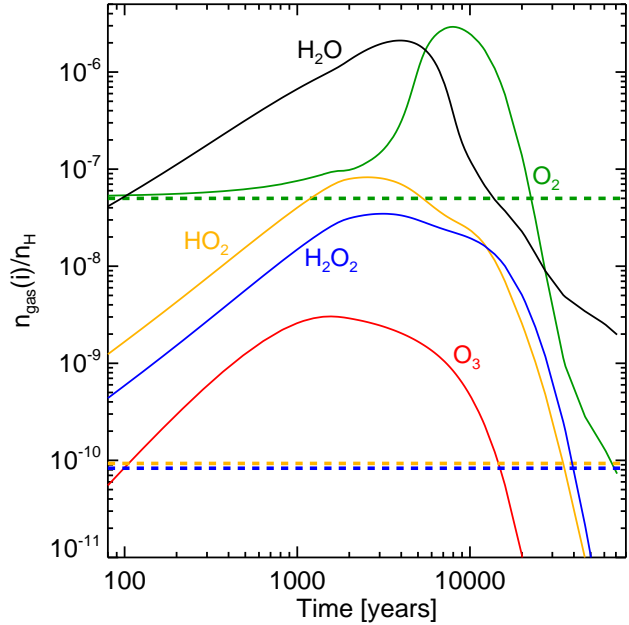} 
\caption{Gas phase abundances of \ce{O2} and its chemically related
  species as a function of time predicted by the model using the
  $\rho$ Oph A physical conditions and the chemical parameters derived
  in the Appendix.
}
\label{rhoopha_gas}
\end{figure}

Figure \ref{rhoopha_gas} shows the temporal evolution of the gas
phase abundances of \ce{O2}, \ce{O3}, \ce{HO2}, and \ce{H2O2} when
the theoretical chemical desorption probabilities from
\citet{minissale2016}, considered as our standard values, are
assumed. 
The high chemical desorption probability of the reaction O~+~O (68 \%)
allows for an efficient evaporation of \ce{O2} in the gas phase upon
formation on ices, inducing maximal abundances of a few $10^{-6}$
obtained at 8000 years. At longer timescales, the \ce{O2} production
on ices is limited and the gas phase abundance of \ce{O2} decreases
sharply in a few $10^4$ yr due to its efficient freeze-out induced at
the high density $n_{\rm H} = 10^6$ cm$^{-3}$.
The surface reactions \ce{O2}~+~H and \ce{HO2}~+~H forming \ce{HO2} and
\ce{H2O2} have a lower chemical desorption probability of 1.4 and 0.5
\% respectively. These values are nevertheless high enough to produce
gaseous abundances of \ce{HO2} and \ce{H2O2} larger than $10^{-8}$. 
As a consequence, the model fails to simultaneously reproduce the
gaseous abundances of \ce{O2}, \ce{HO2}, and \ce{H2O2} derived in $\rho$ Oph A
since the predicted \ce{HO2} and \ce{H2O2} abundances are higher than
the observations by one order of magnitude when the predicted
\ce{O2} abundance reaches the observed value of $5 \times 10^{-8}$ at
a time of $1.8 \times 10^4$ years. Instead, their abundances are fit
at a slightly longer time of $3\times 10^4$ years.

\begin{figure}
\centering 
\includegraphics[width=\columnwidth]{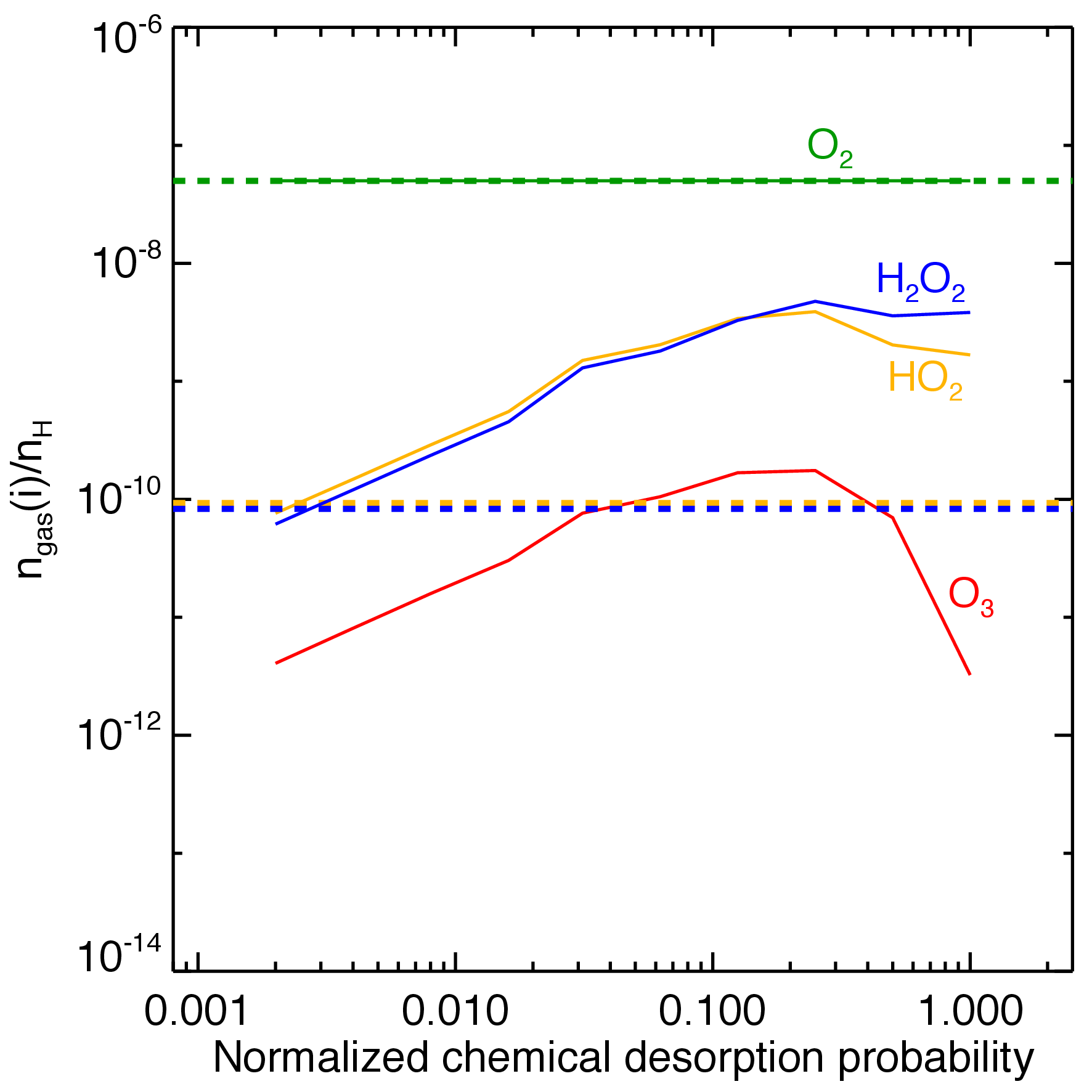} 
\caption{Gas phase abundances of \ce{HO2}, \ce{H2O2}, and \ce{O3}
  obtained when the predicted abundance of \ce{O2} reaches the
  abundance observed at $t = 4 \times 10^4$ yr toward $\rho$ Oph A
  when decreasing the chemical desorption probability of all reactions
  with respect to their standard theoretical value.
}
\label{rhoopha_gas2}
\end{figure}

\citet{du2012} also performed a comprehensive modelling of the gas-ice
chemistry occurring for the physical conditions found in $\rho$ Oph A
by focusing on \ce{HO2} and \ce{H2O2}. Their chemical network is
similar to that used in this work but they used a two-phase model
where the entire bulk ice is assumed to be chemically reactive, and
adopted a high chemical desorption probability of 10 \% for all
surface reactions. 
Their model therefore predicts a high abundance of gaseous \ce{HO2}
and \ce{H2O2}, typically higher than $10^{-8}$ for the first $10^5$ yr
of their simulation, and finds good agreement with the observations,
with abundances of $\sim 10^{-10}$, at $t = 6 \times 10^5$, which is
10 times longer than the free-fall timescale expected at this
density. At this timescale, the predicted \ce{O2} abundance is one
order of magnitude lower than the observed value of $5 \times
10^{-8}$, a similar result as our standard model.
Figure \ref{rhoopha_gas2} shows the gas phase abundances of \ce{HO2},
\ce{H2O2}, and \ce{O3} obtained when the predicted abundance of
\ce{O2} reaches the abundance observed toward $\rho$ Oph A by
decreasing the chemical desorption probability of all reactions with
respect to their standard theoretical value. The model using a
normalized chemical desorption probability of 1 is the standard
model.
It can be seen that the \ce{O2}, \ce{HO2}, and \ce{H2O2} abundances
can be simultaneously reproduced when chemical desorption
probabilities lower than the standard values by a factor of 500 are used,
giving absolute values of $\sim$ 0.001 \% for the reactions H~+~\ce{O2}
and H~+~\ce{HO2}.

\section{Protostellar or disk formation origin?}

The models presented and discussed in the preceding 
Section show that \ce{O2} (and chemically related species) 
can be efficiently formed under dark cloud conditions, 
reaching abundance levels (relative to water ice) similar to that 
observed in comet 67P/C-G as long as the density is high, the
ionisation rate is low, and the temperature is warm.  

Here, we discuss the role of chemistry during protostellar collapse and 
protoplanetary disk formation 
on the observed abundance of \ce{O2} in 67P/C-G.   
Material en route from the protostellar envelope into the disk 
is subject to increasing temperatures and UV radiation generated by the central 
(proto)star.  
We address the following two questions: 
(i) can \ce{O2} gas and/or ice efficiently form during the formation
of protostars and disks if the material composition is initially 
poor in molecular oxygen? and
(ii) can \ce{O2} ice co-formed with \ce{H2O} ice in the prestellar
stage be delivered to the comet-forming zone in young protoplanetary disks 
without significant chemical processing?
To address those questions, the chemical evolution from prestellar
cores to forming disks is calculated.  

\subsection{Model description}

For the protostellar disk formation model, 
the axisymmetric semi-analytical two-dimensional model developed by
\citet{visser2009,visser2011} and adjusted by \citet{harsono2013} is adopted.
Briefly, the model describes the temporal evolution of the density and 
velocity fields following inside-out collapse and the formation of an accretion disk 
described by the $\alpha$-viscosity prescription 
\citep{shakura1973,lynden-bell1974,shu1977,cassen1981,terebey1984}.  
Additional details can be found in the original papers.  
The vertical structure of the disk is calculated assuming hydrostatic equilibrium.
The dust temperature and UV radiation field, which are critical for the chemistry, 
are calculated at each time step by solving the radiative transfer with 
RADMC-3D\footnote{http://www.ita.uni-heidelberg.de/\~dullemond/software/radmc-3d/}.
Outflow cavities are included by hand in a time-dependent manner 
\citep[see][for details]{drozdovskaya2014}.
Initially, the core has a power-law density distribution 
$\propto r^{-2}$, 
where $r$ is the distance from the center of the core,
with an outer boundary of $\sim7000$~AU and a total mass of 1 $M_{\odot}$.
Two values for the initial core rotation rate are investigated: 
$\Omega = 10^{-14}$ s$^{-1}$ and $10^{-13}$ s$^{-1}$,
corresponding to cases 3 and 7 in \citet{visser2009}, respectively.
The model follows the physical evolution until the end of the main 
accretion phase when the gas accretion from the envelope 
onto the star-disk system is almost complete.

Fluid parcels from the envelope to the disk are traced in the physical model, 
and the Furuya astrochemical model is used to follow the gas-ice
chemical evolution calculated along each individual trajectory with
the parameters described in Section \ref{astrochem}. 

A molecular cloud formation model is run to determine the 
composition of the gas and ice in the parent molecular cloud 
\citep{furuya2015}.  
The chemistry is then evolved for an additional $3 \times 10^5$ yr 
under prestellar core conditions to compute the abundances at the onset of 
collapse.  
The prestellar core density, temperature, and visual extinction are set to 
$4\times10^4$~cm$^{-3}$, 10~K, and 10~mag, respectively.
At the onset of collapse, most oxygen ($\gtrsim 95$\%) is contained in  
icy molecules, e.g., \ce{H2O} and CO ice.
The \ce{O2} gas and ice abundances with respect to hydrogen nuclei 
are only $3\times10^{-8}$ and $\ll 10^{-14}$, respectively,
while the \ce{H2O} gas and ice abundances are $2\times10^{-8}$ and 
10$^{-4}$, respectively.  
Hence, the models using this set of initial abundances have a negligible \ce{O2} ice 
abundance.  
Note that the \ce{O2} gas abundance in both the molecular cloud formation stage 
and the prestellar core stage is lower than a few $\times10^{-8}$, which is consistent with 
the upper limits of the observationally derived \ce{O2} gas abundance 
toward nearby cold ($T \sim 10$ K) clouds \citep{goldsmith2000,pagani2003,furuya2015}.  

Following on from the preceding Section, we also explore whether \ce{O2} 
ice co-formed with \ce{H2O} ice in the 
prestellar stage can be delivered to the comet-forming midplanes of protoplanetary 
disks without significant alteration.   
To do this, we also run models with an artificially increased 
initial \ce{O2} ice abundance, set to be 5\% of that for \ce{H2O} ice.

\subsection{Results}

\begin{figure*}
\centering
\includegraphics[width=\textwidth]{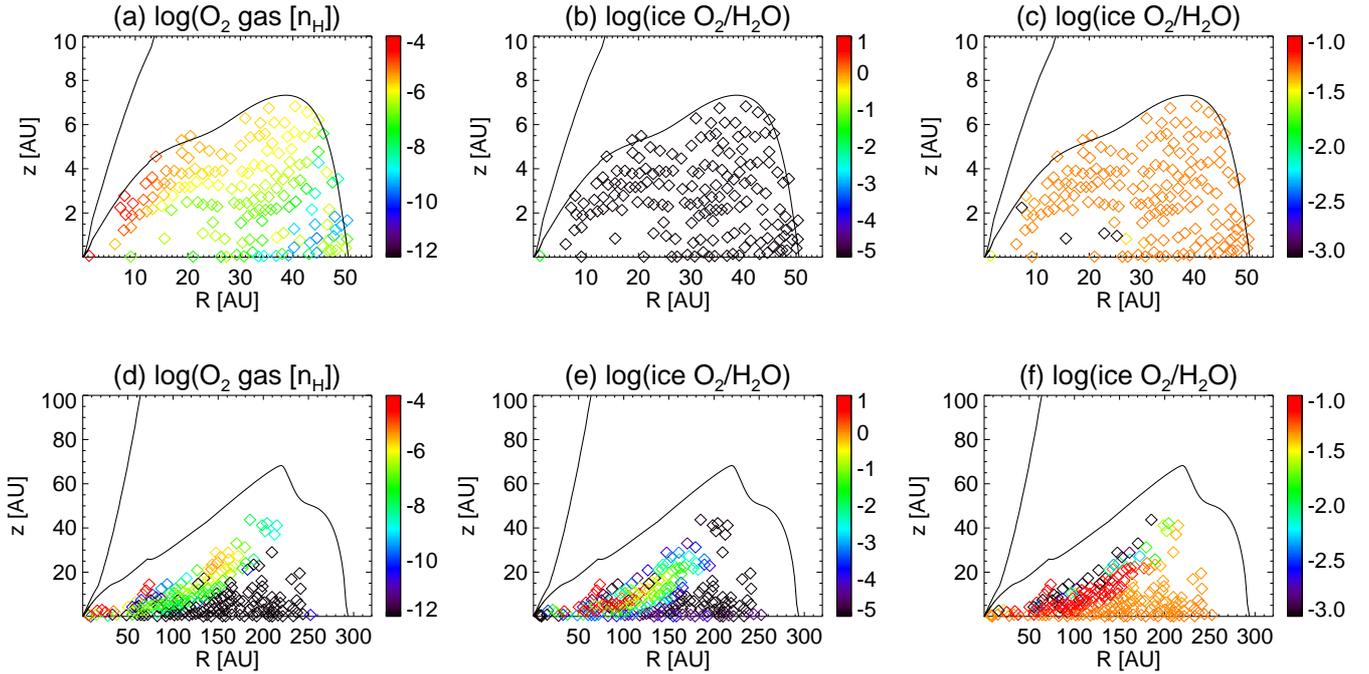}
\caption{Spatial distributions of fluid parcels at the final time of the simulation.
The top panels (a, b, c) represent the collapse model with $\Omega=10^{-14}$ s$^{-1}$, 
while the bottom panels (d, e, f) represent the model with $\Omega=10^{-13}$ s$^{-1}$.
The left panels (a, d) show the gaseous \ce{O2} abundance with respect to hydrogen nuclei,
while the middle panels (b, e) show the abundance ratio between \ce{O2} ice and \ce{H2O} ice.
The right panels (c, f) also show the abundance ratio between \ce{O2} ice and \ce{H2O} ice,
but for those models where the initial ratio is artificially set to 5\%.
The solid lines represent the outflow cavity wall and the disk surface.}
\label{fig:disk_formation1}
\end{figure*}

Figure \ref{fig:disk_formation1} shows the spatial distributions of 
fluid parcels at the final time of the simulation 
in models with $\Omega = 10^{-14}$ s$^{-1}$ (infall dominated, top panels) 
and $10^{-13}$ s$^{-1}$ (spread dominated, lower panels).
For the case in which the ice mantle is poor in \ce{O2} ice 
at the onset of collapse, it is found that
(i) some gaseous \ce{O2} can form (up to $\sim$10$^{-6}$) 
depending on the trajectory paths (left panels), and 
(ii) \ce{O2} ice trapped within \ce{H2O} ice does not efficiently form en route into 
the disk (middle panels).

\begin{figure}
\centering
\includegraphics[width=80mm]{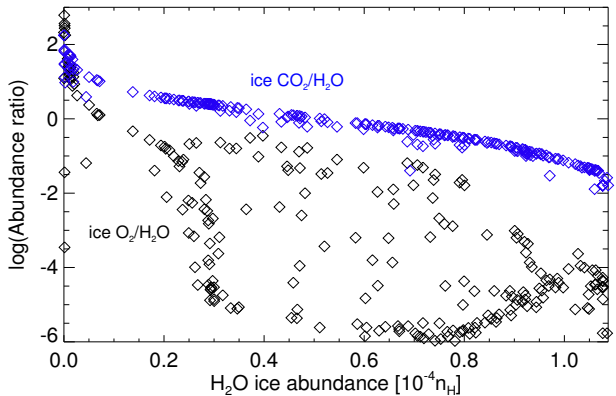}
\caption{\ce{O2} ice (black) and \ce{CO2} ice (blue) abundances
  relative to \ce{H2O} as a function of \ce{H2O} ice abundance at the
  final time of the simulation in the model with $\Omega=10^{-13}$
  s$^{-1}$ and the very low initial \ce{O2} ice abundance.  }
\label{fig:disk_formation2}
\end{figure}

Given that most elemental oxygen is in ices (\ce{H2O} and \ce{CO}) 
at the onset of collapse, gaseous \ce{O2} forms
through photodissociation/desorption of \ce{H2O} ice by stellar UV photons 
in the warm ($>$20 K) protostellar envelope, followed by 
subsequent gas-phase reactions (e.g., O~+~OH).
The middle panels of Figure~\ref{fig:disk_formation1} show that
the majority of parcels in each disk have a low 
final \ce{O2}/\ce{H2O} ice ratio, $\ll 10^{-2}$.  
However, the upper layers of the larger (i.e, higher $\Omega$ case) disk 
do have several parcels with a \ce{O2}/\ce{H2O} ice 
ratio higher than $10^{-2}$ (see panel (e) in Figure~\ref{fig:disk_formation1}).
Analysis of the ice composition shows that the 
\ce{O2} ice is associated with \ce{CO2} ice rather than with \ce{H2O}.   
Upon water ice photodissociation, the warm temperatures encountered through 
the protostellar envelope mean that  
\ce{CO2} ice (re)formation is more favorable than that for \ce{H2O} ice.  
This is due to the weak binding energy of atomic hydrogen: 
the reaction to form \ce{CO2} ice (via, e.g., CO~+~OH) proceeds faster 
that that for \ce{H2O} reformation (e.g., H~+~OH) as atomic hydrogen 
escapes back into the gas phase before it can diffuse and react with OH.  
Figure \ref{fig:disk_formation2} shows the correlation among 
the abundances of \ce{H2O} ice, \ce{O2} ice, and \ce{CO2} ice in the 
model with $\Omega = 10^{-13}$ s$^{-1}$.
In regions where \ce{O2} ice is relatively abundant ($>$1 \% of \ce{H2O} ice), 
the \ce{CO2} ice abundance is higher than 
or comparable to the \ce{H2O} ice abundance. 
Hence, these results show that it is difficult to form \ce{O2} ice
which is closely associated with \ce{H2O} ice during the process 
of core collapse and disk formation.     

For the case that the simulations begin with an appreciable 
fraction of \ce{O2} ice (5\% relative to water ice), 
the \ce{O2}/\ce{H2O} ratio throughout both disks is largely preserved.  
This is indicated by the relatively homogenous distribution of orange points 
in panels (c) and (f) in Figure~\ref{fig:disk_formation1}.  
Hence, \ce{O2} which has a prestellar or molecular cloud origin, 
is able to survive the chemical processing en route into the comet-forming 
regions of protoplanetary disks.  
Trajectories which are an exception to this rule, are those which 
have been most exposed to stellar radiation; however, these 
trajectories are predominantly in the upper and closer-in layers of 
each protoplanetary disk and likely do not contribute to the composition of 
the comet-building material.  
This is consistent with the earlier finding by \citet{visser2011} that 
most water ice is delivered to protoplanetary disks without alteration 
or sublimation.

\section{{O$_2$} formation and trapping in disks induced by luminosity outbursts?}

\subsection{Motivation}

The simulations in the previous Section show that \ce{O2} 
can be produced in the gas phase in the intermediate layers of
relatively warm forming disks (see panel (a) in Figure~\ref{fig:disk_formation1}), 
with an abundance a few percent that 
of water ice (i.e., a fractional abundance of $\sim 10^{-6}$ with respect to $n_\ce{H}$). 
The origin of the gas-phase \ce{O2} is driven by photoprocessing of water ice 
by stellar UV photons en route into the disk, which releases 
photofragments required for forming \ce{O2} (O and OH) into the gas-phase.  
Relatively high abundances of gas-phase \ce{O2} are also predicted in the 
inner regions of protoplanetary disks around already formed stars 
\citep[e.g.,][]{walsh2014,walsh2015}.  
The origin of gas-phase \ce{O2} in these models is similar to that in forming 
disks, except that the release of photofragments of water ice photodissociation 
occurs over the lifetime of the disk ($\gtrsim 10^6$~yr) and is driven by
the UV photons generated near the disk midplane by the interaction of cosmic
rays with \ce{H2}.  
\ce{O2} persists in the gas-phase near the disk midplane because its
volatility is such that  it cannot freezeout at the midplane
temperatures within a few 10's of AU (typically $>20$~K).

\begin{figure*}
\includegraphics[width=0.45\textwidth]{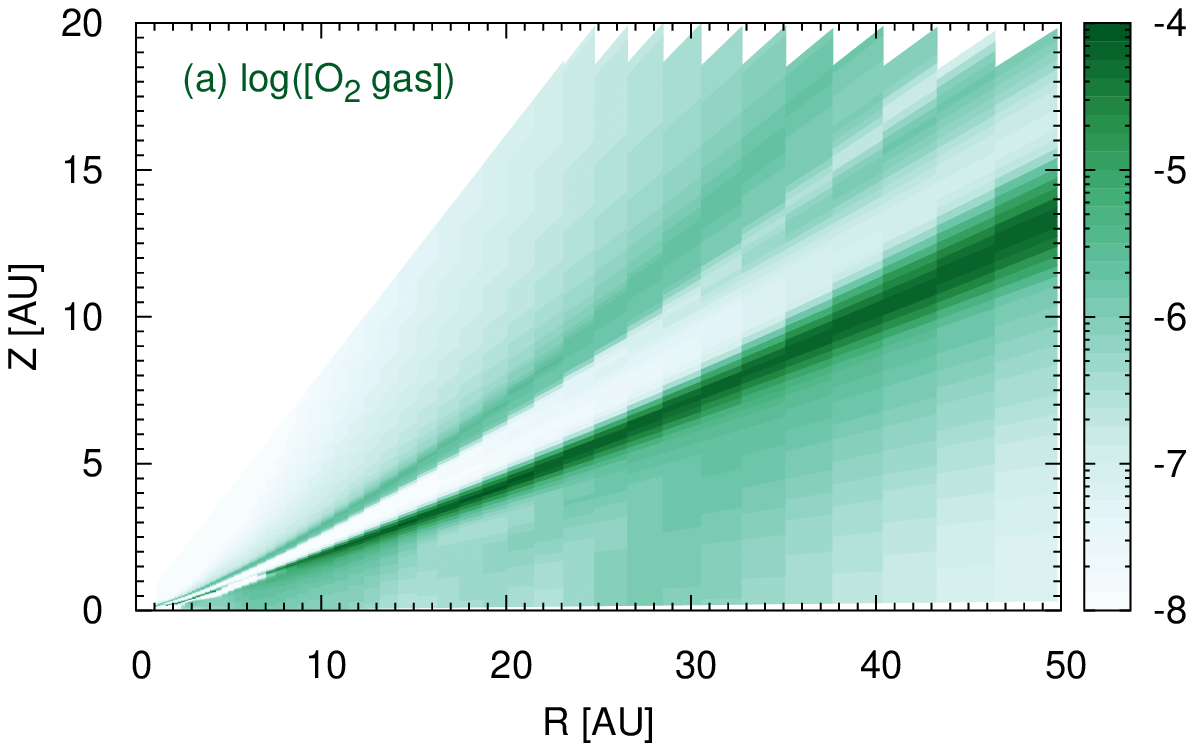}
\includegraphics[width=0.45\textwidth]{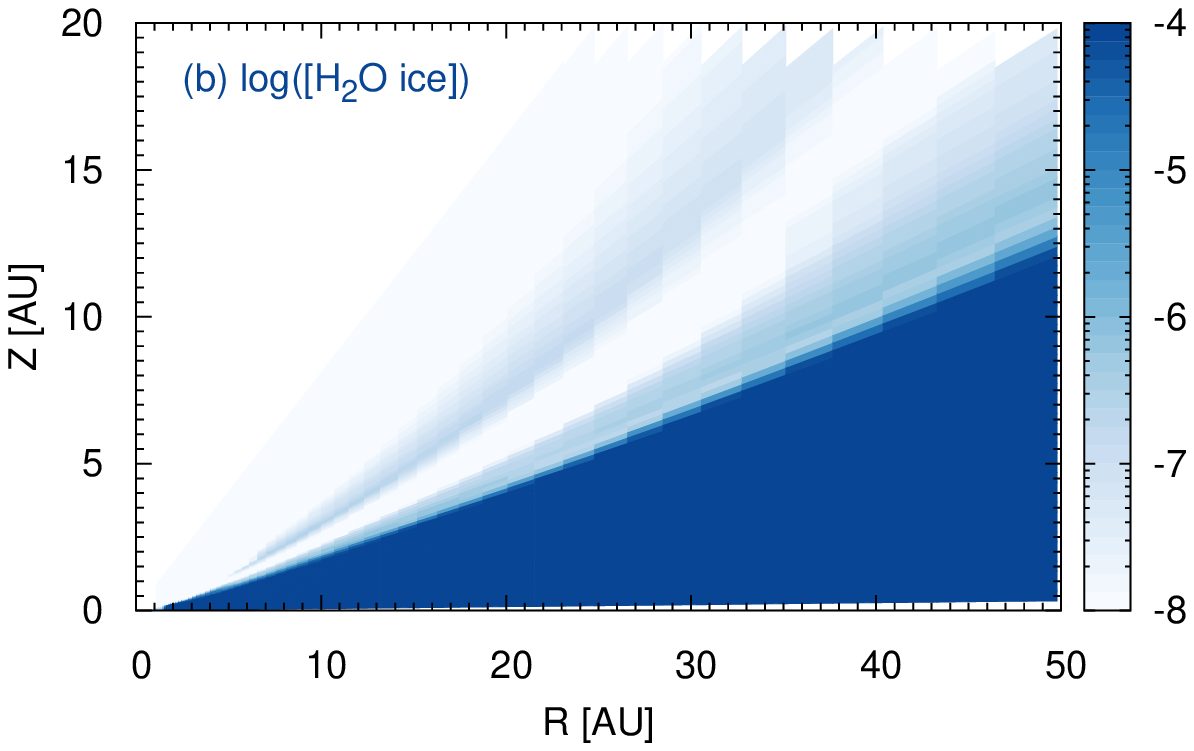}
\caption{Fractional abundance (relative to \ce{H2}) of \ce{O2} gas (left) and \ce{H2O} 
ice (right) as a function of disk radius and height, for a protoplanetary disk around 
a T~Tauri star \citep[data from][]{walsh2014}.}
\label{fig:diskO2-H2O}
\end{figure*}

Figure~\ref{fig:diskO2-H2O} shows the fractional abundance of \ce{O2} gas (left) 
and \ce{H2O} ice (right) as a function of disk radius and height for a 
protoplanetary disk around a T~Tauri star \citep[data from][]{walsh2014}.  
Similar abundances are seen for disks around both cooler (i.e., M~dwarf) and 
hotter stars (i.e., Herbig~Ae) stars, except that the water snowline is shifted 
to smaller and larger radii, respectively \citep[see][]{walsh2015}. 
The results show that \ce{O2} gas can reach an abundance a few percent of that 
of water ice in the comet-formation zone ($\lesssim 50$~AU).  
   
The main issue with this scenario is whether a mechanism exists whereby 
gas-phase \ce{O2} formed near the disk midplane, in either forming
disks or more evolved disks, can become entrapped within, and thus
associated with, the water-rich ice mantle, as seen in comet 67P/C-G.  
Observational and theoretical studies suggest that the luminosity
evolution of low-mass stars is highly variable, with
frequent and strong eruptive bursts, followed by long periods of
relative quiescence \citep[e.g.,][]{herbig1977,hartmann1985,vorobyov2005}.
Such luminosity outbursts could have a strong impact on the morphology
and the chemical composition of ices near the protoplanetary disk
midplane. 
The sudden temperature variations induced by
short luminosity outbursts could gradually recycle the content of ices
into the gas phase and modify their chemical structure via rapid and 
efficient freeze-out.
If the luminosity outburst is sufficiently strong, 
warm gas-phase formation of molecular oxygen could be triggered 
by the evaporation of water ice, if the peak temperature during the outburst 
is higher than $\sim 100$~K. 
\ce{O2} might then be recondensed together with water post outburst, if the
cooling timescale is shorter than the freeze-out timescale 
i.e., $\tau_\mathrm{cool} < \tau_\mathrm{fr}$,
and also if the temperature reached after post outburst is lower than the 
condensation temperature of \ce{O2} ($\approx20$~K). 

An increase in temperature from $\approx 20$ to $\approx 100$~K 
during an outburst, 
corresponds roughly to an increase in luminosity by a factor of $\sim 600$
assuming that the temperature in the disk and the central luminosity
are linked through Stefan-Boltzmann's law.
The recent hydrodynamical model by \citet{vorobyov2015}
shows that a dozen of such strong luminosity outbursts, with typical
durations of $10-100$~yr,  may occur during the disk lifetime.  
The exact number depends on the physical properties of the
collapsing core and the disk.

\subsection{Model description}

The scenario of formation and recondensation of
\ce{O2} induced by a series of outburst events in disks is
investigated by a series of outbursts occurring
every $10^4$ yr for a total timescale of $10^5$ yr.
The astrochemical model and chemical network used are described in 
Section \ref{astrochem}, while the assumed physical conditions are for
a single point, motivated by protoplanetary disk models. 
Initial ice abundances are the median values derived by \citet{oberg2011} 
from interstellar ice observations towards low-mass protostars. 
Thus it is assumed that the ice mantles are initially poor in \ce{O2}.  
The pre-outburst and post-outburst temperature is set to 20~K,
corresponding approximately to the freeze-out temperature of \ce{O2}.  
Protoplanetary disk models suggest that the corresponding 
midplane density at this point is $\sim 10^8$ cm$^{-3}$ 
\citep[e.g.,][]{furuya2013,walsh2014}; 
however, the exact relation between the dust temperature and gas density 
near protoplanetary disk midplanes depends on numerous factors 
including disk surface density (or mass), stellar spectral type, and the dust 
properties.   

Gas phase formation of \ce{O2} is triggered by the photodissociation
of water into H and OH and consequently, is highly dependent on the assumed 
cosmic-ray ionisation rate, $\zeta$, which is thought to be impeded 
near the disk midplane with respect to interstellar values \citep[e.g.,][]{cleeves2013}. 
The impact of $\zeta$ on the formation of \ce{O2} is investigated by
considering two values which cover the  possible range, $\zeta = 1
\times 10^{-18}$ and $1 \times 10^{-17}$ s$^{-1}$. 

The freeze-out timescale of a neutral species $i$ onto grains is given by
\begin{align}
\tau_{\textrm{fr}} = & 1.6 \times 10^2 \textrm{yr} \, \frac{10^8\textrm{cm}^{-3}}{n_{\textrm{H}}} \, 
	\frac{10^{-2}}{R_{\textrm{dg}}} \, \times \nonumber  \\
	& \qquad \frac{\rho_{\textrm{d}}}{3 \textrm{g/cm}^{-3}} \, 
        \frac{a_{\textrm{d}}}{1 \mu{\textrm{m}}} \sqrt{\frac{10 \textrm{K}}{T}} 
        \, \sqrt{M_{\textrm{i}}}, 
\label{freezeout}
\end{align}
where $R_{\textrm{dg}}$ is the dust-to-gass mass ratio, $\rho_{\textrm{d}}$
the volumic mass of grains, $a_{\textrm{d}}$ the mean grain
diameter, and $M_{\textrm{i}}$ the weight of species $i$. 
Grain growth is expected to occur near protoplanetary disk
midplanes. 
\citet{vasyunin2011} predict an average size of 1 $\mu$m
with a dust-to-gass mass ratio of 0.01 in the midplane but the average
size sharply decreases with altitude. 
Therefore, two grain sizes are considered $a_{\textrm{d}} = 0.1$ and 1~$\mu$m.
For a fixed dust-to-gas mass ratio, ${R}_\mathrm{dg}$, a larger grain 
size will increase the freezeout timescale, $\tau_\mathrm{fr}$, 
relative to the cooling timescale, $\tau_\mathrm{cool}$, 
due to the reduction in total available dust-grain surface area.  
On the other hand, an increase in ${R}_\mathrm{dg}$, 
perhaps due to settling and/or radial drift, will increase the 
total available grain surface area and will reduce the freezeout 
timescale.   

Six models are run to investigate the impact of various
parameters on the formation and the recondensation of \ce{O2} during
outbursts which last a timescale, $\tau$. 
\begin{enumerate}
\item Standard model with $a_{\textrm{d}} = 1$ $\mu$m, $\zeta = 1 \times 10^{-18}$ s$^{-1}$,
$\tau = 100$ yr, $T_{\textrm{max}} = 100$ K. 
\item Same as 1 but with $a_{\textrm{d}} = 0.1$ $\mu$m. 
\item Same as 1 but with $\zeta = 1 \times 10^{-17}$ s$^{-1}$. 
\item Same as 1 but with $\tau = 10$ yr. 
\item Same as 1 but with $T_{\textrm{max}} = 150$ K. 
\item Same as 1 but with an initial \ce{O2} abundance of 5 \% relative to water. 
\end{enumerate}

Model~6 is included to test the hypothesis that primordial \ce{O2}, formed 
during the molecular cloud stage, survives both transport into the 
the forming protoplanetary disk and luminosity outbursts 
in the disk midplane.  
For the standard set of parameters (i.e., model 1), 
the freezeout timescale following a burst is 
$100-200$~yr (see Equation~\ref{freezeout}).  
This is likely longer than the cooling timescale 
(from $\approx$~100 to 20~K), expected to be shorter than the duration 
of the outburst \citep[$< 100$~yr;][]{vorobyov2015}.

\subsection{Results}

\begin{figure*}
\centering
\includegraphics[width=0.7\textwidth]{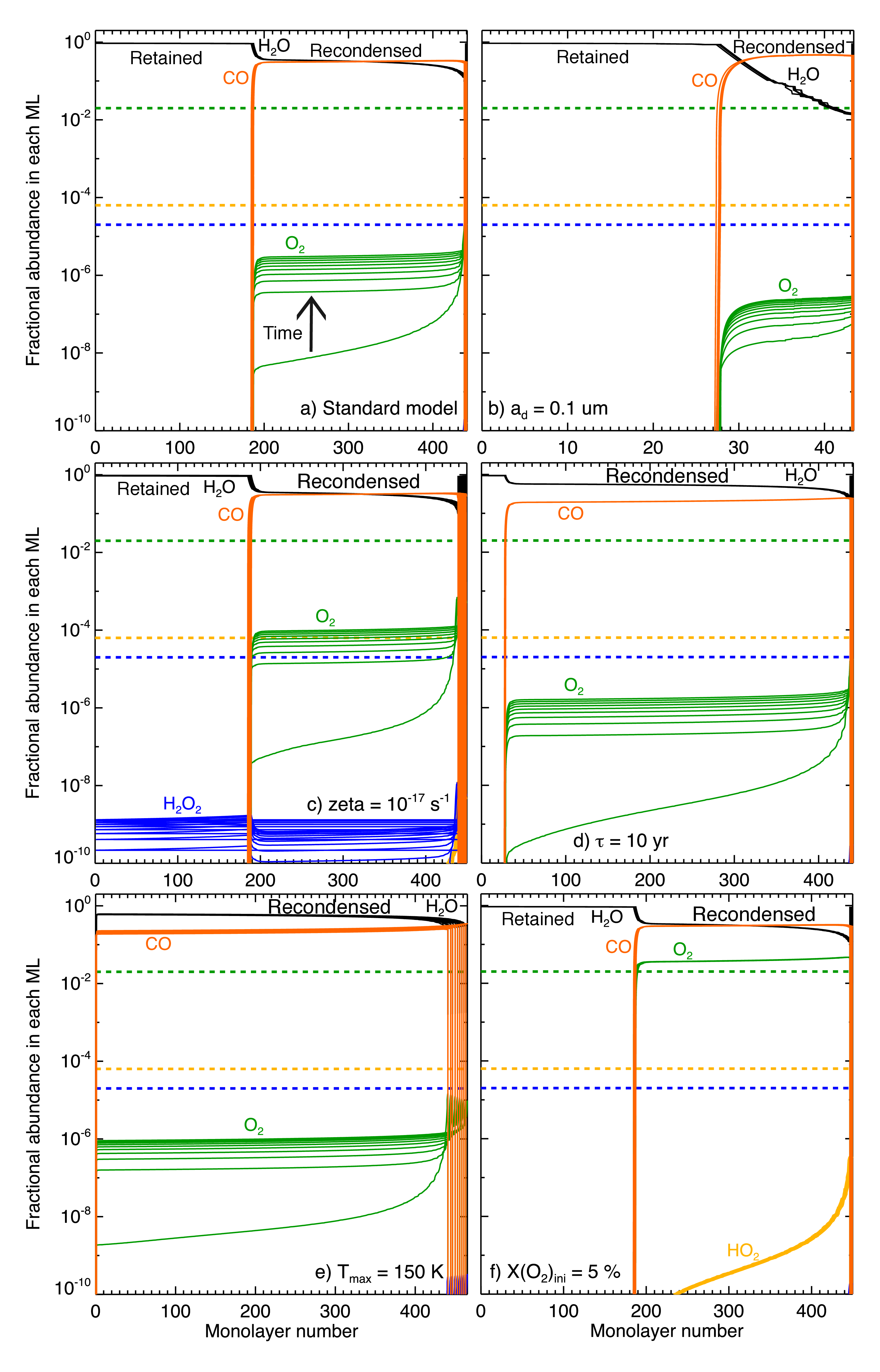}
\caption{Fractional composition of each monolayer within 
ices during the 10 luminosity outbursts for the six models 
considered in this work (see the text for more details). 
The standard parameters are $a_{\textrm{d}} = 1$~$\mu$m, 
$\zeta = 1 \times 10^{-18}$~s$^{-1}$, 
$\tau = 100$~yr, $T_{\textrm{max}} = 100$~K. 
The thick dashed lines refer to the abundances
observed in the comet 67P/C-G.}
\label{outburst_fig}
\end{figure*}

Figure \ref{outburst_fig} shows the fractional composition of 
ices in each monolayer as a function of monolayer, for
the six models described above.  
Each line represents the abundance of each species prior to the 
next outburst (i.e., following each period of cooling and quiescence).  
In general, regardless of the assumed physical parameters, 
the volatile component of the ice mantle increases with time.  
The ice profile is composed of two main parts: 
i) a volatile-free deeper ice mantle mostly composed of water ice and other non-volatile species, 
such as \ce{CH3OH}, that primarily remain on grains during the outbursts
because the temperature reached during the outburst is only slightly
higher than their evaporation temperature, 
and ii) an upper ice mantle 
composed of water ice, but also of volatile species, such as CO, \ce{O2}, and \ce{N2}, 
that freezeout during the post-outburst cooling. 
The deeper and volatile-free ice mantle increases in mass/depth with the duration
of the outburst (compare panels a) and d) in Figure~\ref{outburst_fig}) 
and with decreasing grain size that increases the surface area of dust 
(compare panels a) and b) in Figure~\ref{outburst_fig}).

The fraction of \ce{O2} trapped in the ice mantle 
increases with the outburst duration,
the grain size, and the cosmic-ray ionisation rate, $\zeta$, 
all parameters which favour the
formation of gaseous \ce{O2} from water during the outburst. 
Increasing the grain size decreases the total grain cross-sectional area 
and therefore the accretion rate of gas-phase species onto the ice mantle, 
allowing water and other species to spend more time in the gas phase for reaction. 
A higher $\zeta$ increases the production of OH from the
photodissociation of water vapour, necessary to form \ce{O2}. 
Allowing the peak temperature during outburst to reach values higher 
than the evaporation temperature of water ice 
($T_\mathrm{max} = 150$~K, panel e) in Figure~\ref{outburst_fig}), 
results in full sublimation of the ice mantle prior to recondensation.  
However, the enhanced abundance of water released into the gas phase
does not significantly enhance the abundance of \ce{O2} formed, and 
subsequently trapped. in the ice mantle.  
The luminosity outburst period and duration considered here 
are potentially too short and also too infrequent to reproduce 
the high amount of \ce{O2} observed in 67P/C-G.  
Maximum \ce{O2} abundances of a few $\times 0.1$\% only are predicted.

Results for the calculation with 5\% of \ce{O2} (relative to water ice) 
already present in the ice, show that \ce{O2} 
can survive and be efficiently trapped within the water-rich ice mantle 
following a series of luminosity outbursts.  
Hence, \ce{O2} may become associated with water ice 
in the disk midplane via release and recondensation driven by outbursts.  
However, other volatile species such as CO and \ce{N2} are also trapped 
within the water ice, which is in contradiction with the observations towards 
67P/C-G.  
CO and \ce{N2} are shown to be depleted in 67P/C-G relative to interstellar 
values, and the molecules are not strongly correlated with water in the comet coma, 
converse to the case for \ce{O2} \citep[][]{rubin2015a,bieler2015}.  
Note also that interstellar ices produced after luminosity outbursts are likely
amorphous in structure \citep{kouchi1994}.

\section{Summary}
\label{summary}

In this work, sophisticated astrochemical models are used to investigate 
the chemical and physical origin of molecular oxygen in comet 67P/C-G as
observed with {\it Rosetta}/ROSINA.  
The observations show that molecular oxygen is not only strongly associated with 
water, but is also the fourth most abundant molecule in the coma, 
at $\sim$4\% of the water abundance \citep[][]{bieler2015}.  

We tested the formation and survival of \ce{O2} (and related species) 
in models covering a range of dark cloud physical conditions 
(temperature, density, and cosmic-ray ionisation rate).
We found that the efficiency of the formation of molecular oxygen increases  
for higher densities ($\gtrsim 10^{5}$~cm$^{-3}$), moderate 
temperatures ($\approx 20$~K), and moderate ionisation rates ($\lesssim 10^{-16}$~s$^{-1}$).  
These conditions lower the ratio of H/O in the gas phase, thereby 
impeding the conversion of \ce{O2} ice into \ce{H2O} ice.
These parameters are found to be in good agreement with the physical
conditions for the dark cloud $\rho$~Oph~A, one of the two interstellar
regions where \ce{O2} has been detected. 
The high \ce{O2} abundances do not require photolysis of bulk ice as the
main process but are the result of surface reactions building up the
ice layers. 

We next tested whether molecular oxygen can be efficiently formed in
the ice mantle during protostellar disk formation.  
For models in which the initial ice composition is assumed to 
be poor in \ce{O2}, \ce{O2} can produced only through gas phase
chemistry induced by processing of the water-rich ice mantle by
stellar UV radiation in the intermediate-layers of the protoplanetary
disk, with abundance levels similar to that in 67P/C-G but ices in the
disk midplane remain poor in \ce{O2}.  
For models in which the ice mantles were originally abundant in molecular 
oxygen ($\approx 5$\% relative to water), for both disk models, 
we find that the oxygen is delivered to the comet-forming zone without
sublimation nor alteration.  
Hence, if molecular oxygen has a primordial origin as suggested by the 
dark cloud models, then it can survive transport into the protoplanetary disk.  

Given that gas-phase \ce{O2} can form near protoplanetary disk midplanes, 
and reach abundances relative to water ice similar to that in 67P/C-G, 
we finally tested whether luminosity outbursts which increase the local temperature 
to $> 100$~K, aid the formation and entrapment of gas-phase \ce{O2} 
into the water-rich ice mantle. { Although laboratory experiments
  show that \ce{O2} can be efficiently formed within water ices during
  the ice recondensation through radiolysis \citep{teolis2006},
  we consider this less likely because the cosmic-ray ionisation rate and
  energetic particles from the (pre)solar wind are expected to be
  significantly attenuated near the disk midplane. } 
It is found that the maximum amount of \ce{O2} formed during
luminosity outbursts and then trapped within the ice mantle during the
cooling depends on several parameters, such as grain size, ionisation
rate, or the outburst duration, but never exceeds $\sim 0.1$\%.  
Assuming an initial \ce{O2} abundance of $5$\% relative to water ice
results in an efficient trapping of \ce{O2} within the water-ice
mantle due to the fast cooling after the outburst. 
However, in that case also other volatile species, such as CO and \ce{N2}, 
become trapped, which is in contradiction with observations towards
67P/C-G.  

In summary, the models presented here favour the scenario that molecular 
oxygen in 67P/C-G has a primordial origin (i.e., formed in the molecular cloud) 
and has survived transport through the protostellar envelope and into the 
comet-forming regions of protoplanetary disks.
{The ``primordial'' origin of \ce{O2} is in good agreement with
  the conclusions of \citet{mousis2016}. However, while
  \citet{mousis2016} invoked radiolysis to efficiently convert water
  ice to \ce{O2}, we find that the entrapment and strong association
with water ice combined with low abundance of species like \ce{H2O2},
\ce{HO2}, or\ce{O3} can { alternatively} be explained by an
efficient \ce{O2} formation at the surface of interstellar ices
through oxygen atom recombination in relatively warmer ($\sim 20$ K)
and denser ($n_{\textrm{H}} \gtrsim 10^5$ cm$^{-3}$) conditions than
usually expected in dark clouds.} 
The { weak correlation} of CO and \ce{N2} with
water seen in 67P/C-G is explained by a later formation of these
species in dark clouds with respect to \ce{O2} and water. 
This picture would therefore be consistent with the physical and chemical
properties of our Solar System, such as the presence of short-lived radio isotopes in
meteorites or the orbits of Solar System planets, suggesting that our
Solar System was born in a dense cluster of stars \citep[see][]{adams2010}.

\section*{Acknowledgements}

The authors thank T. Lamberts, E. Bergin, and the ROSINA team,
especially K. Altwegg, M. Rubin, and A. Bieler, for fruitful
discussions and comments on the manuscript and M. Persson for making Figure 1.
Astrochemistry in Leiden is supported by the European Union A-ERC
grant 291141 CHEMPLAN, by the Netherlands Research School for
Astronomy (NOVA), by a Royal Netherlands Academy of Arts and Sciences
(KNAW) professor prize.
K.F. is supported by the Research Fellowship from the Japan Society
for the Promotion of Science (JSPS). 
C.~W.~acknowledges support from the Netherlands Organization for
Scientific Research (NWO, program 639.041.335). 








\appendix

\section{Impact of chemical parameters on the composition of interstellar ices}

A set of models is run, in order to investigate the impact of various
surface and chemical parameters on the chemical composition of interstellar ices
and to assess whether the abundances of \ce{O2}, \ce{O3}, \ce{HO2},
and \ce{H2O2} observed in comet 67P/C-G can be reproduced simultaneously. 
In each model, the ``standard'' values of the input parameters, listed
in Table \ref{grid_table}, are assumed and only one of the parameters
is varied in turn. 
In particular, the physical conditions assumed here
are conditions that favour a high production of \ce{O2}, i.e. a high
density $n_{\textrm{H}} = 1 \times 10^6$ cm$^{-3}$ and a warm
temperature $T = 20$ K, according to the discussion in Section
\ref{imp_params}. 

\begin{figure}
\centering 
\includegraphics[width=0.33\textwidth]{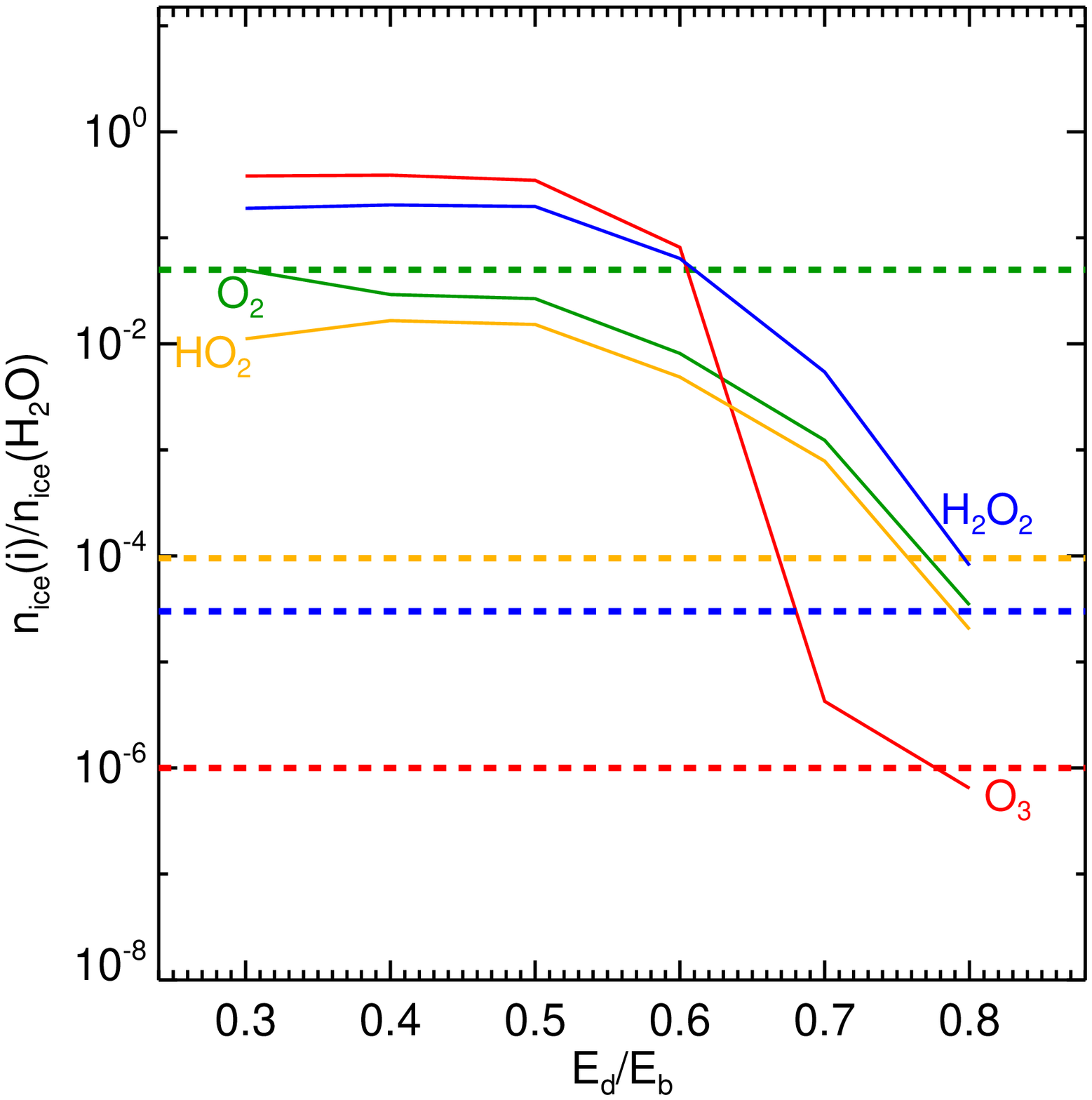}
\includegraphics[width=0.33\textwidth]{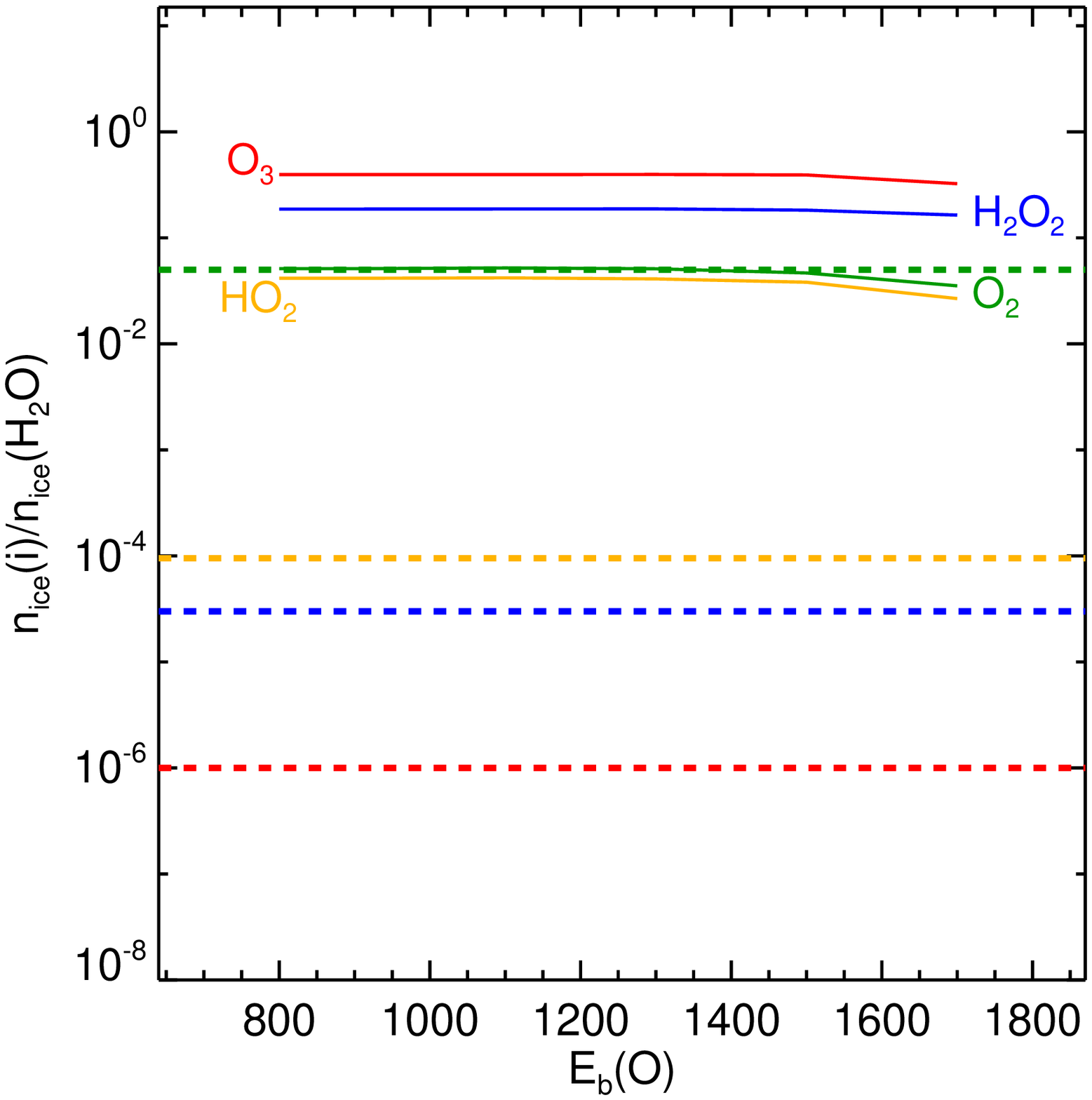}
\caption{Abundances of \ce{O2}, \ce{O3}, \ce{HO2}, and \ce{H2O2} in
  interstellar ices (with respect to water ice) at a 
time of $4.4\times 10^4$ yr, when the gas phase \ce{O2} abundance of
the standard model reaches the abundance observed in $\rho$ Oph A, for
different values of the surface parameters $E_d/E_b$ (left) and
$E_b$(O) (right). 
The thick dashed lines refer to the abundances observed in the comet 67P/C-G.} 
\label{appendix_fig1}
\end{figure}

The impact of two surface parameters, the diffusion-to-binding energy
ratio, $E_d/E_b$, and the binding energy for \ce{O} on the formation
and survival of \ce{O2} ice and the chemically related species is
investigated first.
Following the discussion in Section \ref{imp_params}, models in which
the diffusion-to-binding  energy ratio ranges between 0.3 and 0.8 and
for which the binding energy  of atomic oxygen ranges between 800 and
1700~K have been run. 
Figure~\ref{appendix_fig1} shows the abundances of \ce{O2}, \ce{O3},
\ce{HO2}, and \ce{H2O2} in the solid phase (relative to water ice),   
at a time of $4.4\times10^4$ yr for the different values of the
input chemical parameters. 
The abundance of \ce{O2} (and that for chemically related species) 
tends to decrease as $E_d/E_b$ is increased, both in
the gas phase and in the ice mantle.
The formation rate of \ce{O2} is governed by the mobility of O atoms. 
Due to their relatively high binding energy ($1700$~K for the standard model), 
O atoms can diffuse efficiently only if $E_d/E_b \lesssim 0.6$. 
Higher values strongly inhibit the mobility, leading to a low
abundance of \ce{O2}, \ce{O3}, \ce{HO2}, and \ce{H2O2} (see left panel
in Figure ~\ref{appendix_fig1}).
Decreasing the binding energy of O atoms from 1700 to 800~K 
increases their mobility but also increases the rate of evaporation at 20~K.  
This then limits the conversion from O to \ce{O2} and \ce{O3}, 
leading to a negligible dependence of the final abundances upon the 
assumed binding energy for atomic oxygen 
(see right panel in Fig.~\ref{appendix_fig1}).
The high abundance of \ce{O2} seen in 67P/C-G can be reproduced for dense
and warm conditions and assuming the standard values for the
diffusion-to-binding energy ratio and the binding 
energy of atomic O. 
However, the abundances of \ce{O3}, \ce{HO2},
\ce{H2O2} are overproduced by more than two orders of magnitude when
standard values for the activation barriers of surface reactions are
assumed. 

\begin{figure*}
\centering 
\includegraphics[width=0.33\textwidth]{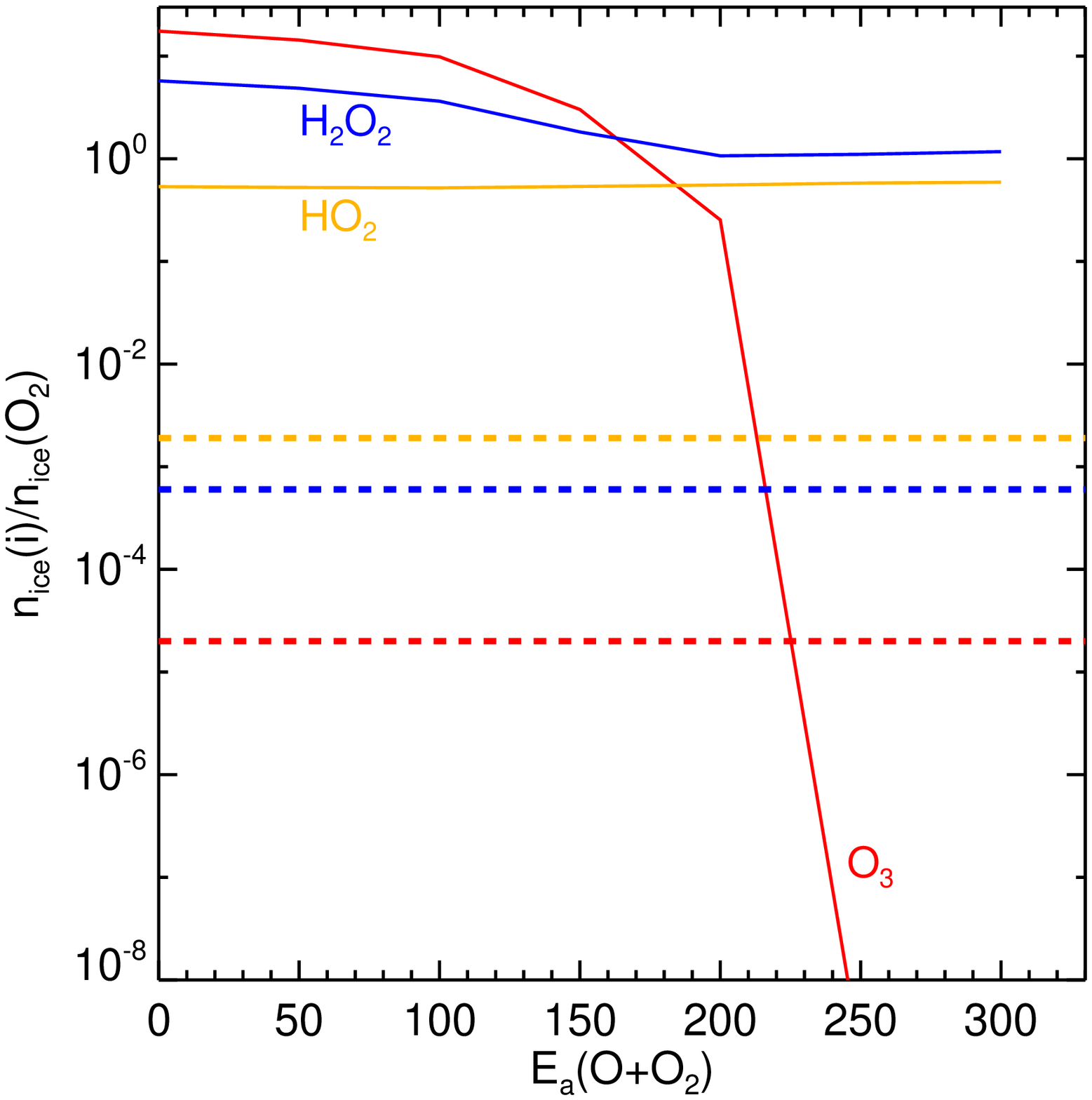} 
\includegraphics[width=0.33\textwidth]{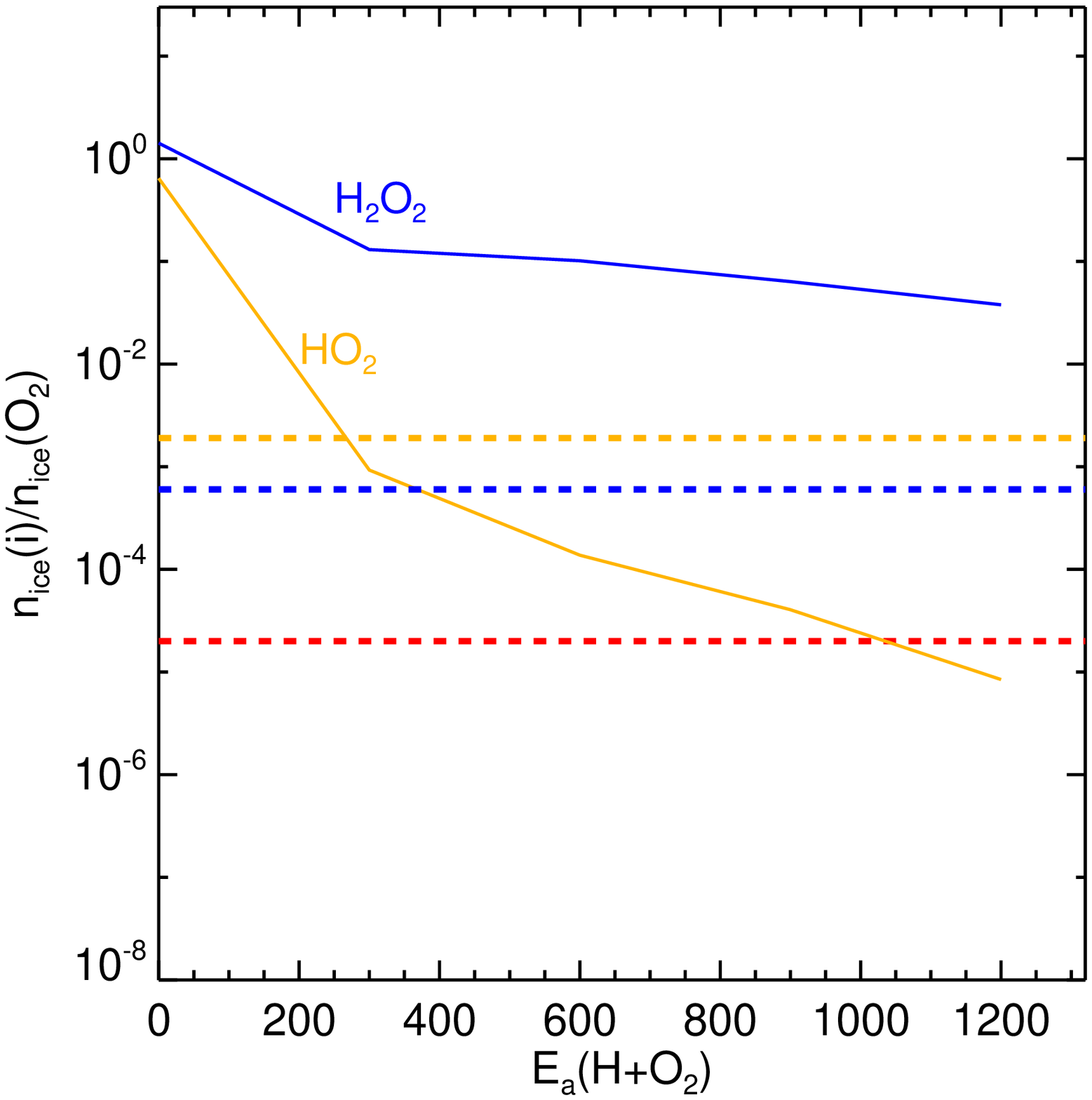} 
\includegraphics[width=0.33\textwidth]{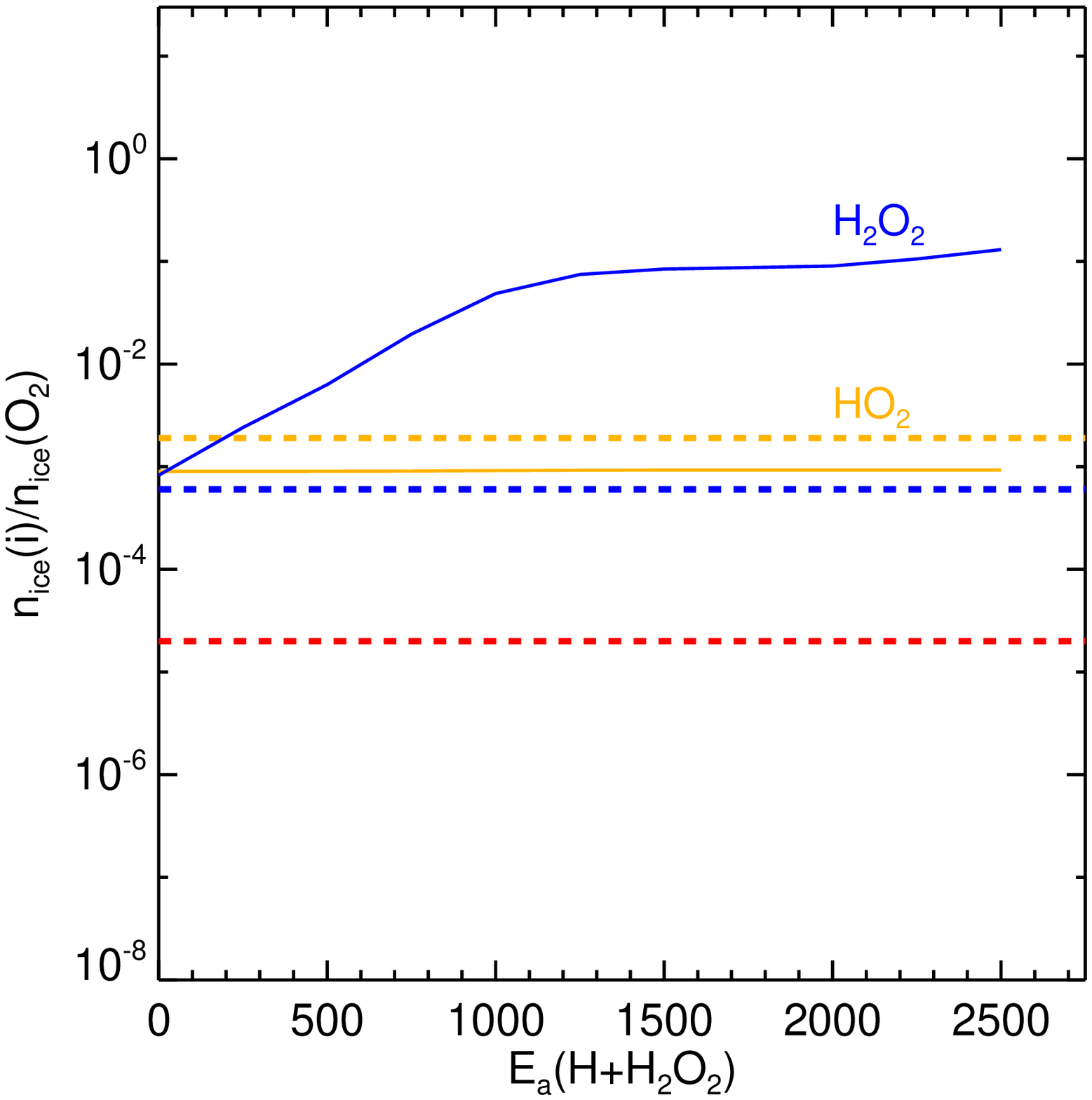}
\caption{Abundances of \ce{O3}, \ce{HO2}, and \ce{H2O2} in interstellar
  ices (with respect to\ce{O2}) at a time of $4.4 \times 10^4$ yr for
different values of the activation barriers of the reactions O +
\ce{O2} (left), H + \ce{O2} (middle, and H + \ce{H2O2} (right). 
Models of the left panels assume the standard values for the
parameters listed in Table \ref{grid_table}, while models of the
middle panels assume an activation barrier for the reaction O +
\ce{O2} of 300 K, and models of the right panels assume an activation
barrier of 300 K for the reactions O + \ce{O2} and H + \ce{O2} (see
text for more details). 
The thick dashed lines refer to the abundances observed in in the
comet 67P. } 
\label{appendix_fig2}
\end{figure*}

The impact of the activation barriers of the reactions O~+~\ce{O2}, H
+ \ce{O2}, and H + \ce{H2O2} on the abundances of \ce{O3}, \ce{HO2},
and \ce{H2O2} in ices is explored next. 
Figure \ref{appendix_fig2} shows how the abundances of \ce{O3}, \ce{HO2},
and \ce{H2O2} in the gas phase (top row) and in the solid phase
(bottom row) relative to \ce{O2} vary with the activation barriers of
these three reactions. 
As described in Section \ref{imp_params}, surface reactions involving 
\ce{O2} have small or negligible reaction barriers but the activation
barrier of the reaction O~+~\ce{O2} remains uncertain. 
Due to the relatively high masses of O and \ce{O2}, 
the activation barrier can be only overcome thermally.     
At a dust temperature of 21~K, the reaction probability 
exponentially decreases from 1 ($E_a=0$~K) 
to $6 \times 10^{-7}$ ($E_a=300$~K). 
As a consequence, a small activation barrier of 250-300 K favours 
the formation and survival of \ce{O2}, \ce{HO2}, and \ce{H2O2} 
with respect to \ce{O3} and allows us to reproduce the low
\ce{O3}/\ce{O2} abundance ratio seen in comet 67P/C-G.
This result is in good agreement with the results of \citet{lamberts2013} who
needed to introduce an activation barrier of 500 K in their
microscopic Monte-Carlo model to explain the slow formation of ozone
observed in laboratory experiments of \citet{ioppolo2010}.
The O~+~\ce{O2} reaction still takes place and induces the 
evaporation of \ce{O3} into the gas phase via chemical desorption,
explaining the high abundance of \ce{O3} in the gas phase.  
However, this occurs at a slower rate than the reactions destroying 
\ce{O3} ice through barrierless surface reactions. 

In the model results, a high abundance of \ce{O2} ice is consistently 
accompanied by similar abundances of \ce{HO2} and \ce{H2O2} ice,
because the H + \ce{O2} reaction is assumed to be barrierless,
following the laboratory experiments of \citet{miyauchi2008} and
\citet{ioppolo2008, ioppolo2010}.
However, quantum chemistry calculations by \citet{walch1991} show
that the reaction in the gas phase has an activation barrier whose 
exact value depends on the incoming angle of the molecule. 
\citet{lamberts2013} introduced a small activation barrier of
200-400 K for this reaction in their Monte-Carlo model to reproduce
the chemical composition observed in cold ices produced in the laboratory
experiments by \citet{ioppolo2010}. 
The activation barrier of the reaction H~+~\ce{O2} was therefore varied between 0
K (our standard value) and 1200 K, the energy computed for gas phase
conditions by \citet{melius1979}. The transmission probability through
quantum tunelling was computed assuming a rectangular barrier
with a standard width of 1 $\AA$. 
As shown in the middle panel of Fig. \ref{appendix_fig2}, the
abundance of \ce{HO2} ice decreases sharply with the activation barrier of
the reaction H~+~\ce{O2}, even for moderate values because its rate of
formation becomes much lower than its rate of destruction while the
decrease of the abundance of solid \ce{H2O2} is more limited. A small
activation barrier of 300 K, similar to the values found by
\citet{lamberts2013}, is therefore sufficient to reproduce the low 
abundance of \ce{HO2} relative to \ce{O2} observed in comet 67P/C-G. 

The standard activation barrier of the H~+~\ce{H2O2} reaction of 2500
K and the associated transmission probability computed with the
Eckart model, follow the quantum chemical calculations for gas phase
conditions presented in \citet{taquet2013}.  
As for other reactions showing high activation barriers, the exact
value of their barrier is highly uncertain and could be lowered in
interstellar ices, due to the van der Waals interactions between the
neighbouring water molecules and the reactants, as shown for instance
by \citet{rimola2014} for the case of the CO + H and \ce{H2CO} + H
reactions. 
\citet{lamberts2013} decreased the activation barrier for the
H~+~\ce{H2O2}  to 800 - 1200 K in their Monte-Carlo model to reproduce
the \ce{H2O2} abundance in laboratory cold ices. 
The right panel of Fig. \ref{appendix_fig2} present the evolution of
the \ce{O3}, \ce{HO2}, and \ce{O3} abundances relative to \ce{O2} for
different values of the activation barrier of the  reaction
H~+~\ce{H2O2}, and assuming an activation barrier of 300 K for the 
reactions O + \ce{O2} and H + \ce{O2}. The model tends to overpredict
the abundance of \ce{H2O2} relative to the abundance observed in 67P,
unless the reaction H + \ce{H2O2} is assumed to be effectively barrierless.

\bsp	
\label{lastpage}
\end{document}